\documentclass[aps,pre,preprint,showpacs,superscriptaddress]{revtex4-1}
\usepackage{epsfig}
\usepackage{natbib}
\usepackage{ulem}
\usepackage{xspace}
\usepackage{xcolor}
\usepackage{bbold}
\usepackage{appendix}
\usepackage{amsmath,amssymb}

\begin{document}

\title{Non-normalizable quasi-equilibrium solution of the Fokker-Planck equation for nonconfining fields }

\author{Celia Anteneodo}
\affiliation{Department of Physics, PUC-Rio, Rio de Janeiro, Brazil}
\affiliation{Institute of Science and Technology for Complex Systems, Rio de Janeiro, Brazil.}
\author{Lucianno  Defaveri}
\affiliation{Department of Physics, PUC-Rio, Rio de Janeiro, Brazil}
\author{Eli Barkai}
\affiliation{Department of Physics, Institute of Nanotechnology and Advanced Materials, Bar-Ilan University, Ramat-Gan
	52900, Israel}
\author{David A. Kessler}
\affiliation{Department of Physics, Institute of Nanotechnology and Advanced Materials, Bar-Ilan University, Ramat-Gan
	52900, Israel}

\begin{abstract}
 We investigate the overdamped Langevin motion 
for particles in a potential well that is asymptotically flat. 
When the potential well  is deep compared to temperature, physical observables like the mean square displacement  
are essentially time-independent over a long time interval, the stagnation epoch.  
However the standard Boltzmann-Gibbs (BG) distribution is non-normalizable, given that the usual partition function is divergent. 
For this  regime, we have previously shown that
a regularization of BG statistics allows the prediction of the values of dynamical and thermodynamical observables in the non-normalizable quasi-equilibrium state. 
In this work, based on the eigenfunction expansion  of the time-dependent solution of the associated Fokker-Planck equation  with free boundary conditions, we obtain an approximate time-independent solution of the BG form, valid for times which are long, but still short compared to the exponentially large escape time. 
The escaped particles follow a general free-particle statistics, where the solution is a an error function, 
shifted due to the initial struggle to overcome the potential well. 
With the eigenfunction solution of the Fokker-Planck equation in hand, we show the validity of the regularized BG statistics and how it perfectly describes the time-independent  regime though the quasi-stationary state is non-normalizable. 
\end{abstract}
\maketitle
\section{Introduction}
\label{sec:intro}

A thermal particle subject to a potential $V(x)$ that is confining at short distances but nonconfining otherwise can present 
long-lived quasi-equilibrium states  for sufficiently low 
temperature $T$. 
Over certain timescales, thermodynamic quantities such as the 
free energy $\mathcal{F}$, energy $E$ and the entropy $\mathcal{S}$, as well as dynamical ones such 
as the mean square displacement (MSD), attain  
a  time-independent value and the virial theorem approximately holds, which immediately raises the question about the possibility of using  Boltzmann-Gibbs statistics~\cite{previous2020}. 
 This is nontrivial because the expression for the equilibrium probability, for instance in one dimension, defined as 
$P_{eq}(x) =  \frac{1}{Z} \;e^{-V(x)/(k_B T) }$, 
where  $T$ is the temperature and $k_B$ is the Boltzmann constant, fails due to the diverging of  the normalizing partition function in the denominator,   $Z = \int_{-\infty}^{\infty} e^{-V(x)/(k_B T) }  \,dx$, for non-confining fields~\cite{Fermi,Plastino}.  However, certain aspects of standard statistical physics and thermodynamics can still be applied through a 
suitable regularization of $P_{eq}(x)$ and observable averages,  as we have 
shown  in previous work~\cite{previous2020}, where 
we developed a general formalism for the problem. This was based on scaling solutions
of the Fokker-Planck equation (FPE) for the probability density function (PDF) $P(x,t)$, and alternatively on finite-box solutions. 
In this paper, we address the problem through the time-dependent solution 
expressed as an eigenfunction expansion~\cite{freezing2020,logpot2011}, then identifying the non-normalizable
quasi-equilibrium (NQE) regime where these solutions become effectively time-independent.

The remaining of the paper is organized as follows. 
The system under study is defined in Sec.~\ref{sec:system} from the perspective of the FPE. The concept of a non-normalizable quasi-equilibrium, and its characteristic phenomenology, is presented in Sec.~\ref{sec:QE}. 
The derivation of the approximate solution of the FPE in the intermediately long-time limit, 
based on the eigenfunction expansion of the solution,  is shown in Sec.~\ref{sec:timesolutions}. The regularization procedure is described in Sec.~\ref{sec:regularization}.
The implications for quasi-equilibrium are discussed in Sec.~\ref{sec:final}.

\section{The system}
\label{sec:system}

We consider the overdamped dynamics of a Brownian particle in one dimension, governed by the Langevin equation  (LE)~\cite{vankampen,Risken}
\begin{eqnarray}
\gamma \frac{dx}{dt} =  F(x) + \, \sqrt{2 \gamma k_B T}\;\eta(t), 
\end{eqnarray}
where $\gamma$ is the damping coefficient, 
$\eta(t)$ is a Gaussian white noise with zero mean and variance $ \langle \eta(t) \eta(t') \rangle = \ \delta(t-t')$, 
and $F(x) = - \partial_x V(x)$ is the force. 
We assume, crucially, that the  potential $V(x)$  has  a well at  the origin and it is  flat for large $x$. 
Alternatively, we can investigate the associated FPE~\cite{vankampen,Risken} for the probability density function (PDF) $P(x,t)$, namely,  
\begin{equation}
{\partial \over \partial t}  P(x,t) = D \bigg\{ {\partial^2 \over \partial x^2} - {\partial \over \partial x} { F(x) \over k_B  T} \bigg\} P(x,t),
\label{eq:FPE}
\end{equation}
 where  $D=k_B T / \gamma$ is the diffusion coefficient.

Both perspectives yield  in principle  the same results, and the PDF that solves  the FPE is obtained by averaging over trajectories $x_t\equiv x(t)$ that are solutions of the LE, that is
\begin{eqnarray}
P(x,t) = \langle \delta(x - x_t) \rangle_p \,,
\end{eqnarray}
where $\langle \ldots \rangle_p$ represents an average taken over all possible paths $x_t$ and $\delta$ is the Dirac delta function. Other observables, such as the MSD, can be evaluated also as $\langle x^2 (t) \rangle = \langle x_t^2 \rangle_p$.

Note that the Boltzmann-Gibbs (BG) solution
 
 \begin{equation} \label{eq:BG}
   P(x) =  \frac{1}{Z} \;e^{-V(x)/(k_B T) }  
 \end{equation}
 would be a stationary solution of the FPE Eq.~(\ref{eq:FPEscaled}) if the potential were confining. 
 However, this expression  will not work for the cases studied here as it is not normalizable. This is because the diffusion cannot be blocked indefinitely in a potential that is flat at long distances ($V(x) \to 0$ for $x\pm \infty$).  
As a paradigm for this kind of potential, let us consider the families
\begin{equation} 
V_\mu (x) = - {  U_0 \over \big(1  + (x/x_0)^2 \big)^{\frac{\mu}{2}}},
\label{eq:v}
\end{equation}

\begin{equation} 
V_{\kappa,\mu} (x) = - { \frac{U_0 }{2} \, \frac{1 + \cos(\kappa \, x/x_0)}{  \big(1  +(x/x_0)^2 \big)^{\frac{\mu}{2}} }   } ,
\label{eq:vo}
\end{equation}
with $\mu>0$ and $\kappa$ a real parameter.  
Moreover for simplicity, we have assumed even functions. 

\begin{figure}[h!]
\centering
\includegraphics[width=0.48\textwidth]{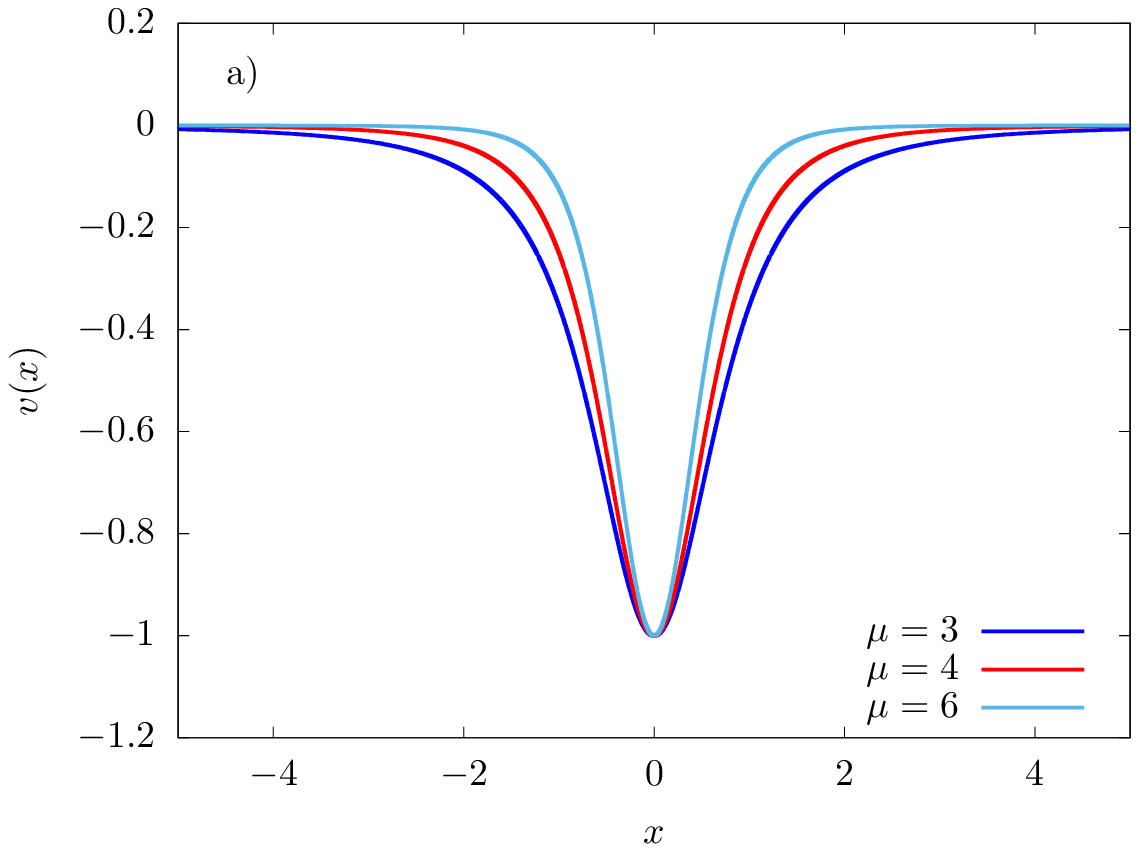}
\includegraphics[width=0.48\textwidth]{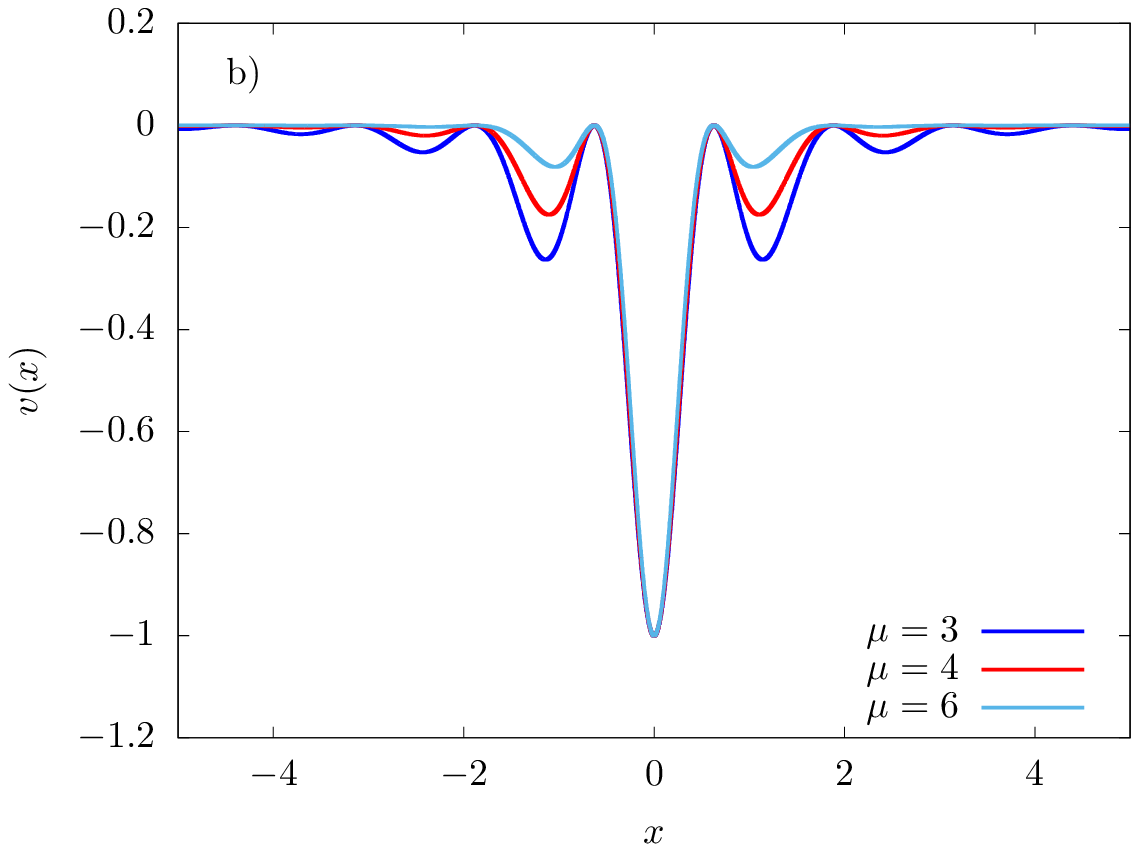}
\includegraphics[width=0.48\textwidth]{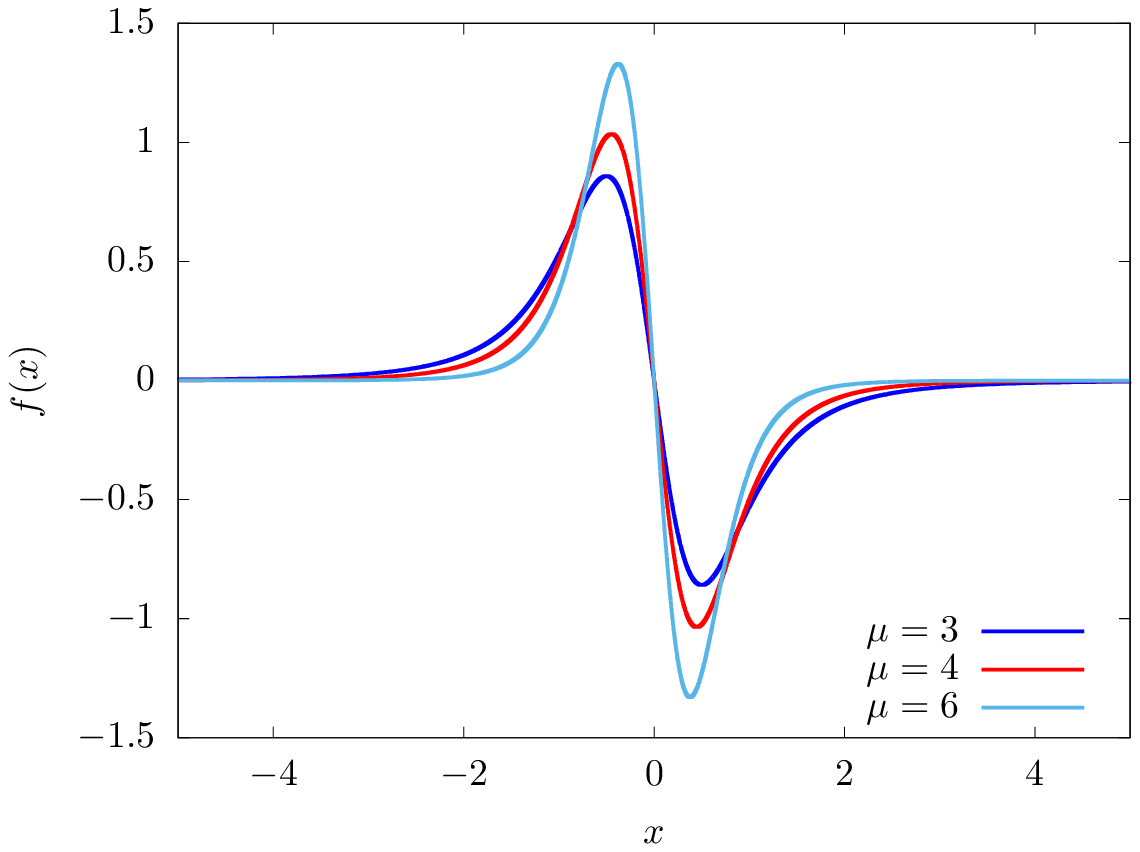}
\includegraphics[width=0.48\textwidth]{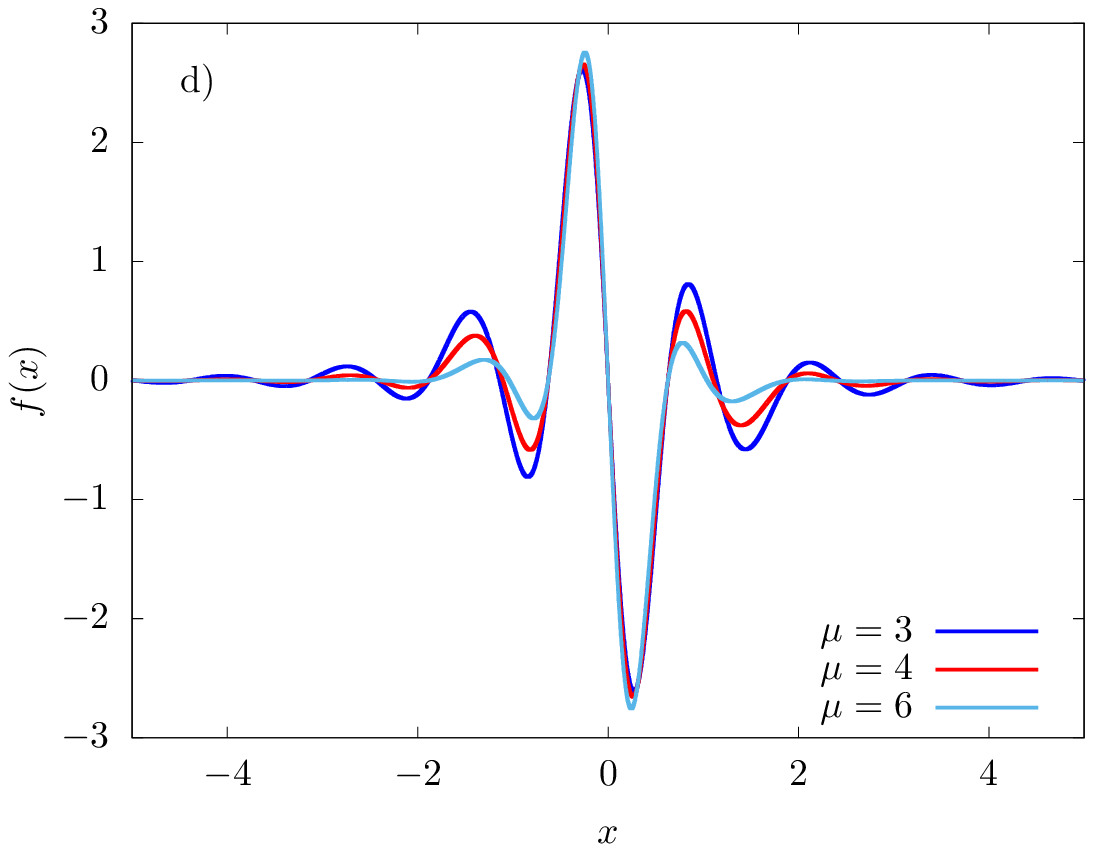}
\caption{
Dimensionless potentials (a) $v_\mu(x)$  and 
(b) $v_{\kappa,\mu}(x)$,  and (c)-(d) the respective forces, plotted for three different values of $\mu$ and $\kappa = 5$. Notice that 
the potential becomes flat and the force falls to zero at large distances from the origin, and therefore ineffective, in the sense that these fields are non-binding and the normalization factor $Z$ in 
Eq.~(\ref{eq:BG}) diverges. 
}
\label{fig:potential}
\end{figure}

It is useful to use dimensionless variables. 
Then we adopt the lengthscale $x_0$ that represents the effective region of the potential well, the timescale  
$t_0 =  x_0^2/D$ related to free diffusion  over this lengthscale, and the energy scale $U_0$ representing the well depth.   
Dimensionless versions of both potential and force can be defined as $v(x) = V(x)/U_0$ and $f(x) = x_0 F(x)/U_0$,  respectively.  A scaled temperature can also be defined as the ratio between the thermal energy and the well depth $\xi = k_B T/U_0$.
After the change of variables  $x/x_0  \to x$,   $t/t_0  \to t$, 
the scaled FPE equation becomes 
\begin{eqnarray}
\frac{\partial}{\partial t} P(x,t) = \frac{\partial^2}{\partial x^2} P(x,t) - \frac{1}{\xi} \frac{\partial}{\partial x} \big( f(x) P(x,t) \big). \label{eq:FPEscaled}
\end{eqnarray}
where the paradigmatic potentials become 
\begin{equation} 
 v_\mu (x) = - { 1 \over \big(1  + x^2 \big)^{\frac{\mu}{2}}},
\label{eq:v}
\end{equation}

\begin{equation} 
 v_{\kappa,\mu} (x) = - { \frac{1}{2} \, \frac{1 + \cos(\kappa \, x)}{  \big(1  + x^2 \big)^{\frac{\mu}{2}} }   } ,
\label{eq:vo}
\end{equation}
and $f(x)=-\partial_x V(x)$.  Notice that $v(0) = -1$.  
The form of these potentials is illustrated in Fig.~\ref{fig:potential} for different values of $\mu$.

\label{sec:PDF}
 \begin{figure}[t!]
\centering
\includegraphics[width=0.48\textwidth]{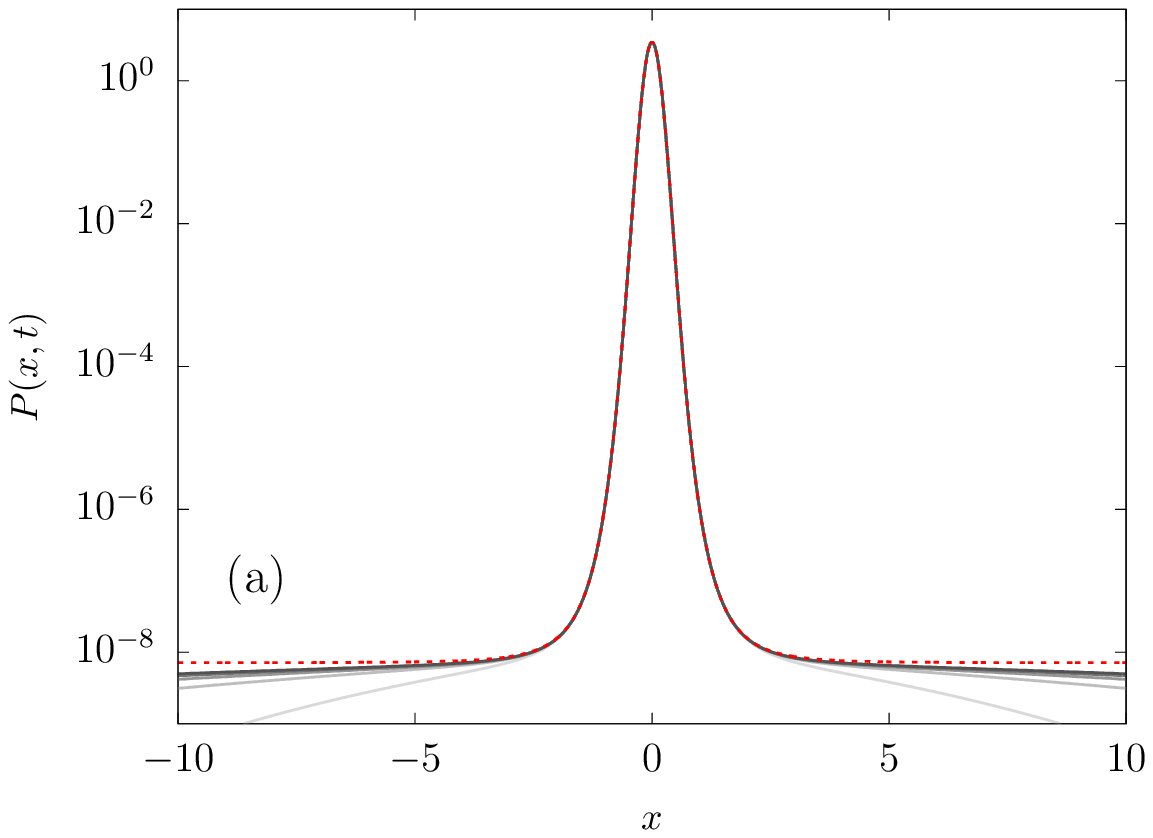}
\includegraphics[width=0.48\textwidth]{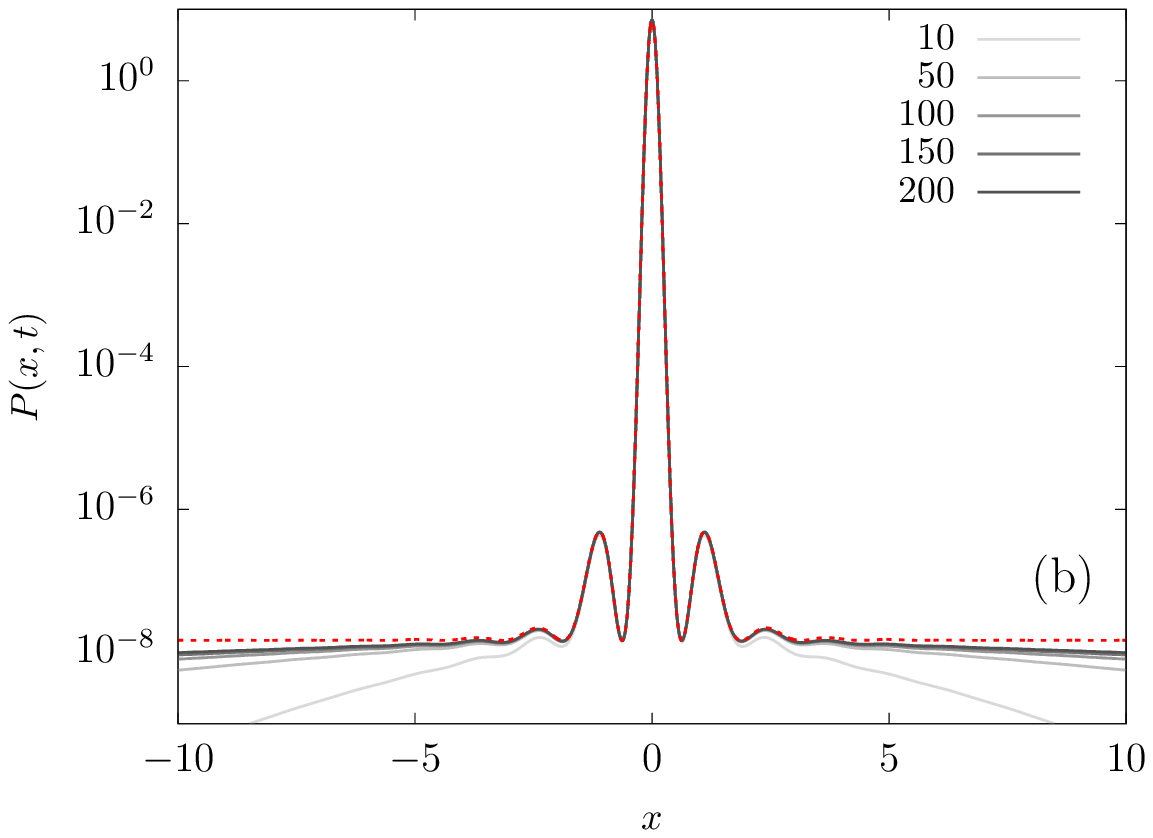}
\caption{
Probability density function $P(x,t)$ for different
times, from the numerical integration of the FPE 
(\ref{eq:FPEscaled})
with the potentials  
(a) $v_4(x)$  and 
(b) $v_{5,4}(x)$,  for $\xi=0.05$. 
For comparison, in each case we also plot (red dotted line) the  Boltzmann expression given by Eq.~(\ref{eq:NQE-BG}), $e^{-v(x)/\xi }/(2\ell_0)$, where $\ell_0$ is defined in  Eq.~(\ref{eq:l0}). 
The maxima in the plots correspond to minima in the potential field.
Note that as we increase time the approximation given by Eq.~(\ref{eq:NQE-BG}) works better, however for times much longer than the escape time a different behavior will be found. 
}
\label{fig:pdf}
\end{figure}

The evolution of a packet of particles can be accessed by numerically 
integrating the FPE  Eq.~\eqref{eq:FPEscaled} in order to obtain the PDF $P(x,t)$. 
The initial condition has the particles starting at the origin, that is $P(x,0) = \delta(x)$. 
The PDF at different times, after a transient, is shown  in Fig.~\ref{fig:pdf} for the potentials $v_4(x)$  and 
$v_{5,4}(x)$. 
Fig.~\ref{fig:pdf} exhibits an interesting  property,  the 
probability density is proportional to the Boltzmann factor 
$exp(- v(x)/\xi)$, i.e., extrema   of the potentials in 
Fig.~\ref{fig:potential} correspond to extrema  of the density in Fig.~\ref{fig:pdf}. Thus, already here we see some concepts of BG statistics are still valid. 
Notice also that most of the probability is in the central  region where the force is significantly non-null. That is, the area under the curve $P(x,t)$ vs $x$ in that central region is nearly one, while outside there are just rare fluctuations.
This motivates the analysis of the FPE instead of 
 finite sample Langevin simulations, as it would require an enormous amount of trajectories to accurately sample that region. 
However, from an experimentalist perspective, the central region is clearly the most important for computing observables of interest.

\section{Non-normalizable quasi-equilibrium}
\label{sec:QE}

 In order to understand the dynamics of the Brownian particle, it is useful to investigate the time evolution of the mean square displacement (MSD), 
\begin{equation} \label{eq:MSDt}
\langle x^2(t) \rangle = \int_{-\infty}^{\infty} x^2 P(x,t) \,dx, 
\end{equation} 
which gives the fluctuation of the position, 
recalling that 
we start the particles on the origin and, from the symmetry of the problem  ($v(x) = v(-x)$), 
$\langle x \rangle = 0$ for all times.

\begin{figure}[h!]
\centering
\includegraphics[width=0.48\textwidth]{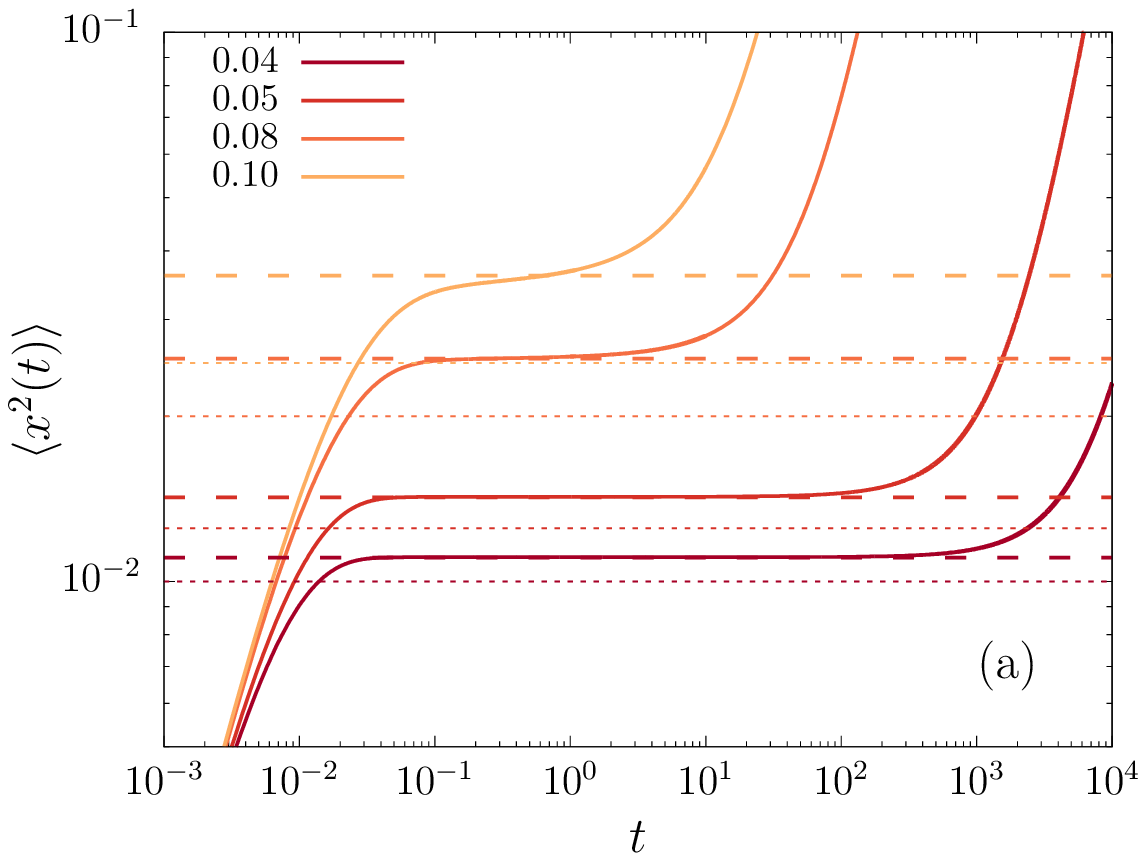}
\includegraphics[width=0.48\textwidth]{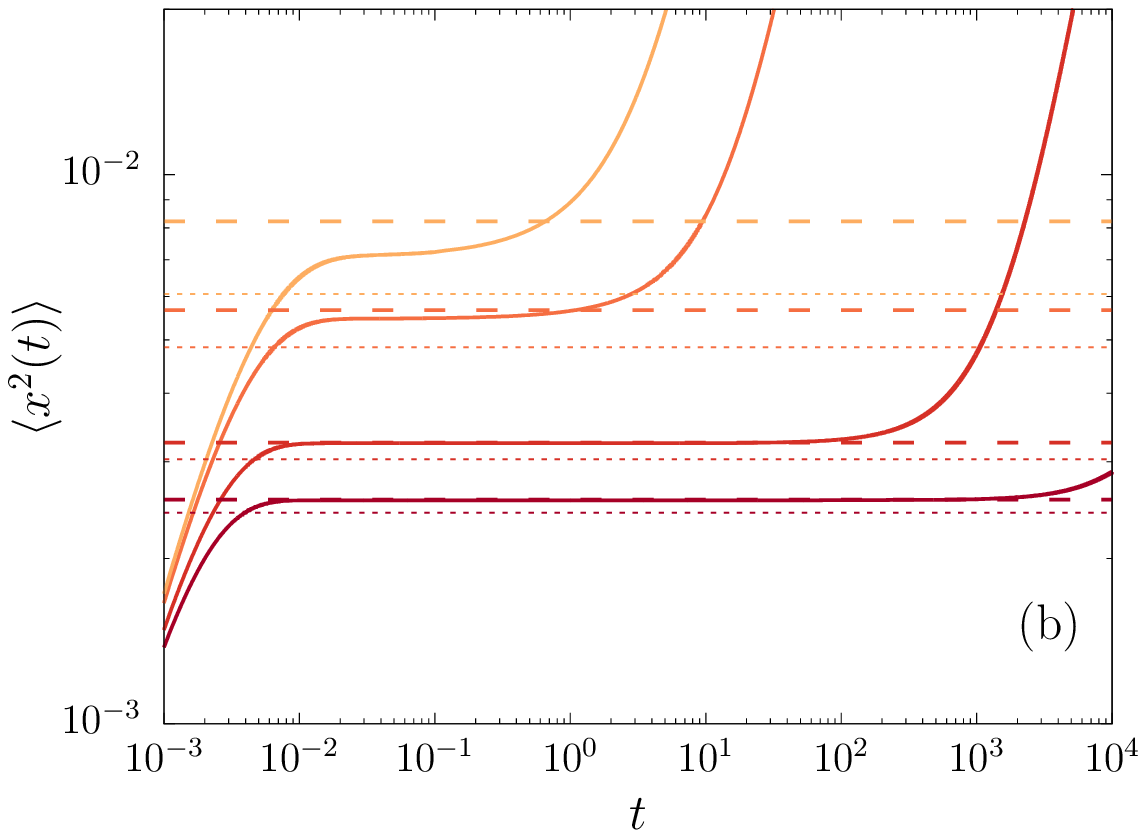}
\caption{
The MSD $\langle x^2 (t) \rangle$  versus time obtained from numerical solutions of the FPE (solid lines) with the potential fields given by 
(a) $v_4(x)$  and 
(b) $v_{5,4}(x)$, and different values of $\xi$ indicated in the legend.
The dashed lines correspond to the expression derived along this work for NQE states, given by Eq.~(\ref{eq:NQEteo}).  
Drawn for comparison, the dotted lines correspond to the harmonic approximation ($\langle x^2 \rangle = \xi/\mu$ and $\langle x^2 \rangle = 2\xi/(\kappa^2 + 2 \mu)$), showing the improvement of the presented theory. 
}
\label{fig:MSDvst}
\end{figure}

The time evolution of the MSD, obtained by numerical integration of the FPE, is illustrated 
in Fig.~\ref{fig:MSDvst}, for the potentials fields in Eqs.~(\ref{eq:v}) and (\ref{eq:vo}) with 
$\mu=4$, at different values of $\xi$.

We observe a short-time increase of the MSD before the particles reach a quasi-equilibrium state (what we called NQE), where the MSD remains almost constant,  yet at even longer times the particles escape the well and normal diffusion leads to an eventually linear increase of the MSD with time. The NQE state is long lived, and, intuitively, the deeper the well depth with respect to temperature the longer is the life time of this stagnated state.
This behavior is observed as well for thermodynamic observables, such as the energy or entropy, as previously shown~\cite{previous2020}.

\begin{figure}[h!]
\centering
\includegraphics[width=0.48\textwidth]{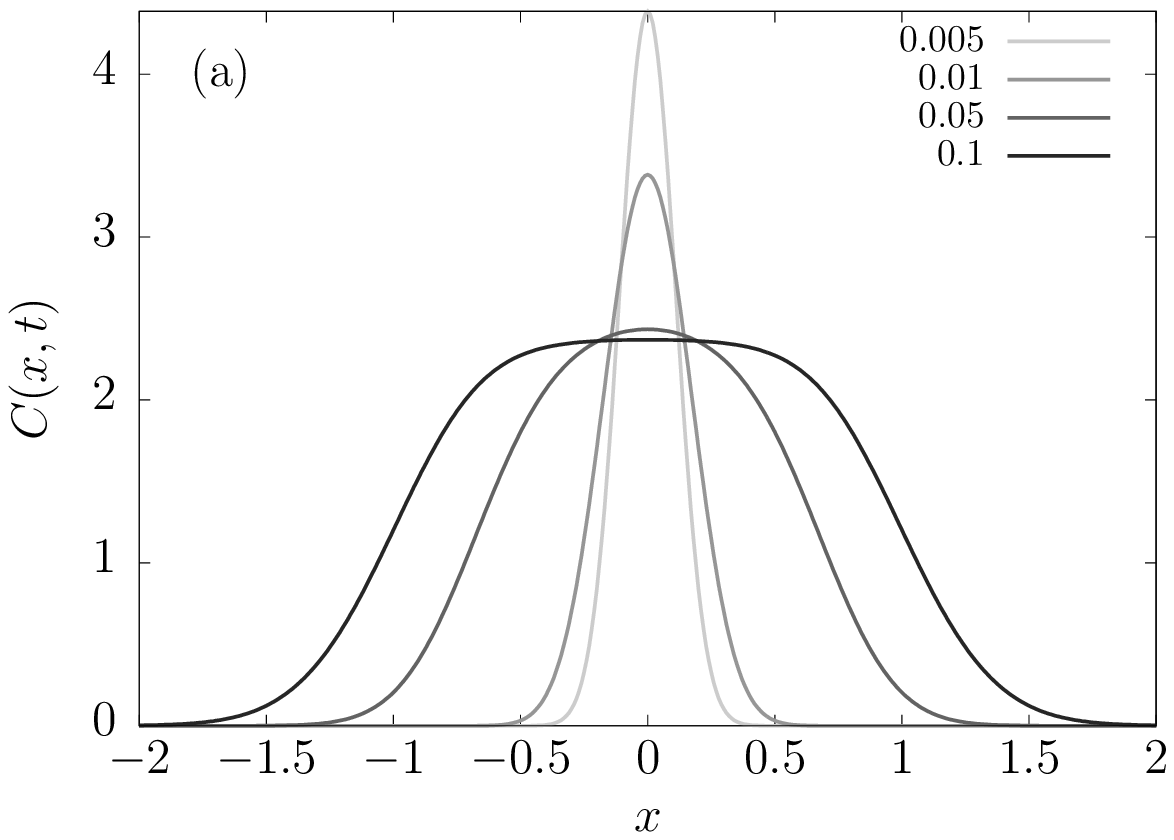}
\includegraphics[width=0.48\textwidth]{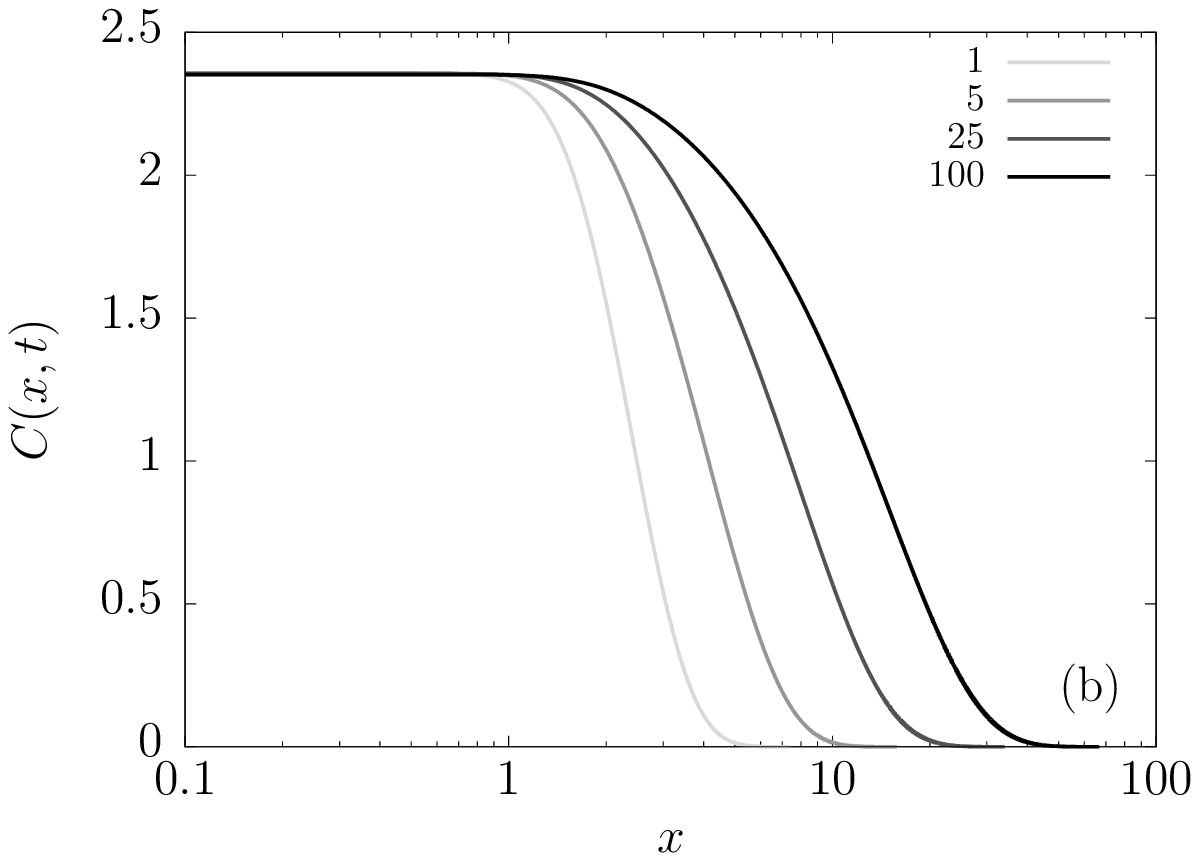} \\
\includegraphics[width=0.48\textwidth]{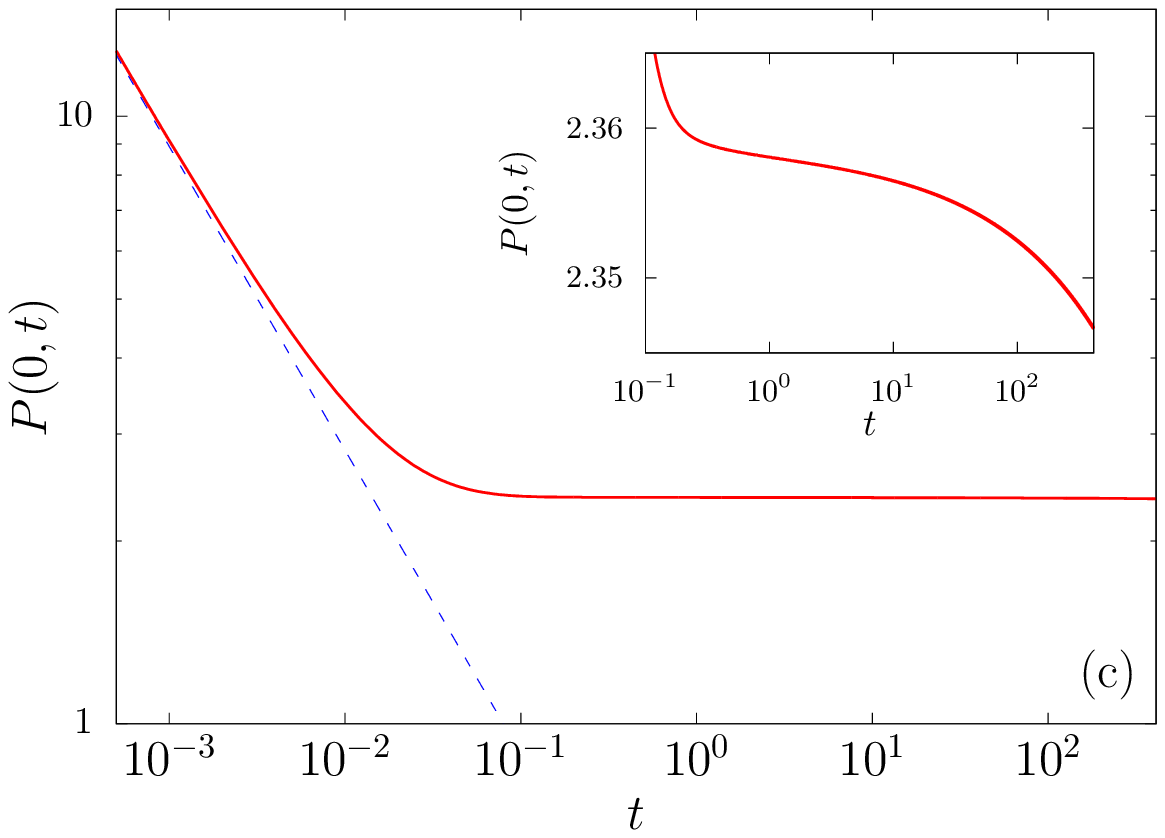}
\caption{
 Pre-factor of the PDF, $C(x,t)$ defined in Eq.~(\ref{eq:Cxt}), 
 at different times,  for the field $v_4$ and effective temperature $\xi=0.1$.  
(a) For very small times, (b) for an intermediate time scale.  
The behavior of the PDF at the origin $P(0,t)$ is shown in (c), with its linear ordinate axis representation in the inset.
}
\label{fig:pdfC}
\end{figure}

To better understand this phenomenon, 
it is useful to write the PDF factoring out the BG factor as 

\begin{equation} \label{eq:Cxt}
    P(x,t)= C(x,t) \, {e}^{-\{v(x)-v(0)\}/\xi} \,,
\end{equation}
where the prefactor $C(x,t)$ is plotted in Fig.~\ref{fig:pdfC}a as a function of $x$ in the upper panels, for different times $t$. 
Notice that there is a large-$x$ cut-off that diffuses away, while the central part flattens 
and attains an almost stationary level (see Fig.~\ref{fig:pdfC}b), namely $C(x,t)$ becomes essentially constant 
in that region after a certain time. 
This means that the PDF approaches the shape defined by Eq.~\eqref{eq:BG}, but the normalization is set by the region where the prefactor is flat. 

 In Fig.~\ref{fig:pdfC}c, we highlight the behavior of $P(x,t)$ at the origin. For small times, we can see  that $P(0,t) \propto 1 / \sqrt{t}$ is a free diffusion until it reaches an approximately constant value,  
while the inset highlights how the value is still decreasing albeit very slowly.
This is valid for times which are long 
compared to the relaxation in the well 
but shorter than the Arrhenius escape time, 
of order $e^{1/\xi}$, what we will call below {\it intermediate times}, but the point is that in experimental situations they can be very long indeed.  
 
Now,   besides the MSD given by Eq.~(\ref{eq:MSDt}), we consider  its time derivative, evaluated as
\begin{eqnarray} \label{eq:derivative}
\frac{d}{dt} \langle x^2(t) \rangle &=& \int x^2 \frac{\partial P(x,t)}{\partial t} \,dx.
\end{eqnarray} 
Using the FPE~\eqref{eq:FPEscaled}, we can expand the expression of the time derivative of the MSD as
\begin{eqnarray}
\frac{d}{dt} \langle x^2(t) \rangle = \int x^2 \frac{\partial P(x,t)}{\partial t} \,dx = \int x^2 \bigg\{ \frac{\partial^2}{\partial x^2} P(x,t) -\frac{1}{\xi} \frac{\partial}{\partial x} \Big( f(x) P(x,t) \Big) \bigg\}\,dx,
\end{eqnarray}
and perform integration by parts to obtain
\begin{eqnarray}
\frac{d}{dt} \langle x^2(t) \rangle = 2 + \big\langle x f(x) \big\rangle = 2 - \left\langle x \frac{\partial v}{\partial x} \right\rangle. \label{msd-v}
\end{eqnarray}

This result allows a qualitative description of the dynamics of a packet  of particles starting at $x=0$, that is, setting $P(x,0) = \delta(x)$.  
(See Fig.~\ref{fig:MSDvst}).

(i) For very short times, the particles spread diffusely, the second term in Eq.~\eqref{msd-v} is null and the force is not yet felt, since by assumption the force vanishes at the origin.

(ii) Once the particles have diffused enough, the influence of the force becomes relevant and slows down the diffusive process. Then, there is a time where the value of the derivative in the left-hand side of Eq.~\eqref{msd-v} becomes minimal, and if the  temperature is small enough ($\xi  \ll 1$),   the system  attains a stationary-like regime, where $d \langle x^2(t) \rangle/ dt \approx 0$, as shown in Fig~\ref{fig:MSDvst} (see also Ref.~\cite{previous2020}). This means that  $\left\langle x  \partial_x v  \right\rangle \approx 2$, which implies that the virial theorem becomes approximately valid.

(iii) Since the force is finite and vanishes for large $x$, it is unable to block the diffusion indefinitely, so after a time that can be exponentially large (that is,  proportional to the Arrhenius factor ${ e}^{ U_0 /k_B T} = { e}^{1/\xi}$), a significant fraction of particles escape to diffuse outside the well. In such case, the second term in Eq.~\eqref{msd-v}  becomes null again, giving rise to  linear growth of the MSD.

\section{Time-dependent solution}
\label{sec:timesolutions}

In this section, we go through the derivation of the time-dependent PDF, $P(x,t)$ for intermediate times,  
over which the prefactor $C(x,t)$ defined in 
Eq.~(\ref{eq:Cxt}) is effectively constant in the central region of the system, as can be seen in Fig.~\ref{fig:pdf}a. 
That is, as commented above, for times which are long 
compared to the relaxation in the well 
but shorter than the Arrhenius escape time~\cite{rednerbook,Kramers,Arrhenius}.  
 
The derivation is structured just as in the non-deep potential case~\cite{AghionPRL,Aghion2020}.
Similar analyses of the time-dependent FPE were   performed in the context of a logarithmic potential~\cite{logpot2011}, 
and front propagation~\cite{sander1998,derrida1997}. 
However, the existence of an intermediate temporal regime is new and its origin needs to be explained. We start from the FPE for the PDF $P(x,t)$, given by Eq.~(\ref{eq:FPEscaled}). 
The observation of a quasi-stationary regime, as depicted in Fig.~\ref{fig:MSDvst}, leads to assume a  
time-independent solution in an intermediate long-timescale. 
Then, we set the left-hand side of the FPE to zero, we obtain the time-independent solution, which we call $\mathcal{I}(x)$,
\begin{eqnarray}
\mathcal{I}(x) = e^{ [v(0) - v(x)]/\xi}. \label{eq:I}
\end{eqnarray} 
This solution, which is Boltzmaniann, satisfies the no-flux boundary condition $[v'(x) I(x)]'=0$. 
However, this solution is not normalizable. 
To circumvent this difficulty, we use a mathematical trick. 
We put the system in a box of size $2L$, where $L$ (measured in units of $x_0$) is much larger than the effective region of the potential well, that is,  $L \gg 1$.  The introduction of these walls at $x=\pm L$ will allow us to normalize the solution. 
On the other hand, heuristically, the particles will diffuse more slowly than a free particle, so we may use the latter case as an upper bound to conclude that as long as $L \gg \sqrt{t}$, the walls are totally irrelevant. In this timescale, our boxed model will be identical to the reality, where the particles  are not limited in space.

The PDF  $P(x,t)$, for the initial condition $P(x,0) = \delta(x)$,  can be written  as the eigenfunction expansion~\cite{Risken},  
\begin{eqnarray} \label{eq:expansion}
P(x,t) = e^{v(0)/\xi} \biggl\{ \mathcal{I}(x) N_0^{-1} + \sum_{\{k\}} N_k^{-1} \Psi_k(0) \Psi_k(x) {  e}^{- k^2 t} \biggr\},
\end{eqnarray}
where $k$ is a wavenumber (scaled by  $1/x_0$) 
given by the no-flux boundary condition at $x=\pm L$, and
$N_k$ is the normalization constant associated to the eigenfucntion $\Psi_k(x)$, with the zero-mode $\Psi_0(x)=\mathcal{I}(x)$. 
Notice that in the limit of large $L$, the eigenvalues spectrum becomes continuous since the potential is non binding,  in essence, this is the same as the spectrum of a free particle. 
This  free particle spectrum has been observed for other potentials too~\cite{freezing2020,logpot2011}. 


We set the normalization of $\Psi_k(x)$ via the condition $\Psi_k(0) = 1$, so that
\begin{eqnarray} \label{eq:N0}
N_0 = 2 \int_0^L \mathcal{I}^2(x) \, { e}^{v(x)/\xi} dx = 2\, {e}^{2 v(0)/\xi} \int_0^L { e}^{-v(x)/\xi}dx,
\end{eqnarray}
 and
 \begin{eqnarray}
 N_k = 2 \int_0^L \Psi_k^2(x) { e}^{v(x)/\xi} dx.
 \end{eqnarray}
Using the FPE, the eigenfunctions $\Psi_k(x)$ satisfy~\cite{Risken}
 \begin{eqnarray}
 \Psi_k''(x) + \frac{1}{\xi}\bigl(v'(x) \Psi_k(x)\bigr)' = -k^2 \Psi_k(x).  \label{eq:SMA3} 
 \end{eqnarray}

The leading zero-mode term of the expansion 
in Eq.~(\ref{eq:expansion}) is simply the Boltzmann steady state in a box $(-L, L)$.
The intermediate-long-time limit is clearly dominated by the small-$k$ modes, since the larger ones are suppressed   as far as $e^{-k^2 t} \ll 1$. 
So, it is enough to consider only the small-$k$ modes.

 We need to treat two regimes separately, first the range $x \ll 1/k$, where the right-hand side of Eq. (\ref{eq:SMA3}) is always small, denoted region {\bf I}, and second, for $x \gg 1$ (region {\bf III}). These two asymptotic limits must be matched in the overlap region $1 \ll x \ll 1/k$ (region {\bf II}).

In region {\bf I}, the term $-k^2 \Psi_k(x)$ is negligible due to the smallness of $k$. To leading order we have the homogeneous equation,
\begin{eqnarray}
\Psi_k''(x) + \frac{1}{\xi} (v'(x) \Psi_k(x))' = 0,
\end{eqnarray}
with the zero-mode solution $ {\cal I}(x)$. To next order, we write $\Psi_k(x) \sim {\cal I}(x) (1 - k^2 g(x))$. Plugging this ansatz into Eq. (\ref{eq:SMA3}), we get
\begin{eqnarray}
- v'(x) g'(x) + g''(x) = 1. \label{eq:g}
\end{eqnarray}  
The boundary conditions translate to $g(0) = g'(0) = 0$, and so a simple calculation yields
\begin{eqnarray}
g(x) = \int_0^x {e}^{v(x_1)/\xi} \left( \int_0^{x_1} { e}^{-v(x_2)/\xi} dx_2 \right) dx_1 \, . \label{eq:g}
\end{eqnarray}
We will soon analyse the large $x$ behavior of $g(x)$ and for that purpose we define 

\begin{equation}
    \label{eq:g1}
g_1(x) \equiv  \int_0^x {e}^{-v(x')/\xi} dx'\,.
\end{equation}
Assuming that $v(x)$ falls fast enough at large $x$ (faster than $1/x$, i.e., $\mu>1$ for the families of potentials we consider), then, for large $x$,
\begin{eqnarray}
g_1(x) &=& \int_0^x \left( { e}^{-v(x)/\xi} - 1 + 1 \right) dx \nonumber \\
			&=& x + \int_0^\infty \left({ e}^{-v(x)/\xi} - 1\right)dx - \int_x^\infty \left({ e}^{-v(x)/\xi} - 1\right)dx \nonumber \\
			&\approx & x + \ell_0\,, \label{eq:ll0}
\end{eqnarray}
where $\ell_0$ is related to the second virial coefficient from the theory of gases~\cite{Montroll}, 
and the integrand is essentially  the Mayer f-function,  namely, 
\begin{eqnarray} \label{eq:l0}
\ell_0 \equiv \int_0^\infty \left( { e}^{-v(x)/\xi} -1 \right)dx \, ,
\end{eqnarray}
which is exponentially large,  of order $e^{1/\xi}$, recalling that 
$v(0)=-1$. 
For instance, for a square well, straightforwardly, $\ell_0 \propto ({e}^{1/\xi}-1)$, for a smooth potential, according to the harmonic approximation $\ell_0 \approx {e}^{1/\xi} \sqrt{\frac{\pi \xi}{2 v''(0)}}$. 
Also, in Eq.~(\ref{eq:N0}), $N_0 \approx 2 \ell_0$, which as we will see will play the role of a regularized partition function. 
Recall that  $v$ falls faster than $1/x$, so that the integral converges. Now, for large $x$,
\begin{eqnarray} \label{eq:g-ap}
g(x) &=& \int_0^x \left( { e}^{v(x_1)/\xi} g_1(x_1) -x_1 - \ell_0 + x_1 + \ell_0  \right) dx_1 \nonumber \\
		&=& \frac{x^2}{2} + \ell_0 x +  \int_0^\infty \left( { e}^{v(x_1)/\xi} g_1(x_1) -x_1 - \ell_0   \right) dx_1 -  \int_x^\infty  \left( { e}^{v(x_1)/\xi} g_1(x_1) -x_1 - \ell_0   \right) dx_1 \nonumber \\
		& \approx & \frac{x^2}{2} + \ell_0 x + {\cal A}\,,
\end{eqnarray}
where    
\begin{eqnarray}
{\cal A} \equiv \int_0^\infty \left( {e}^{v(x)/\xi}g_1(x) - x - \ell_0 \right) dx \, . \label{eq:A}
\end{eqnarray} 
This behavior of $g(x)$ can be seen to be consistent with Eq.~(\ref{eq:g}).

\begin{figure}[b!]
	\centering
	\includegraphics[width = 0.49\textwidth]{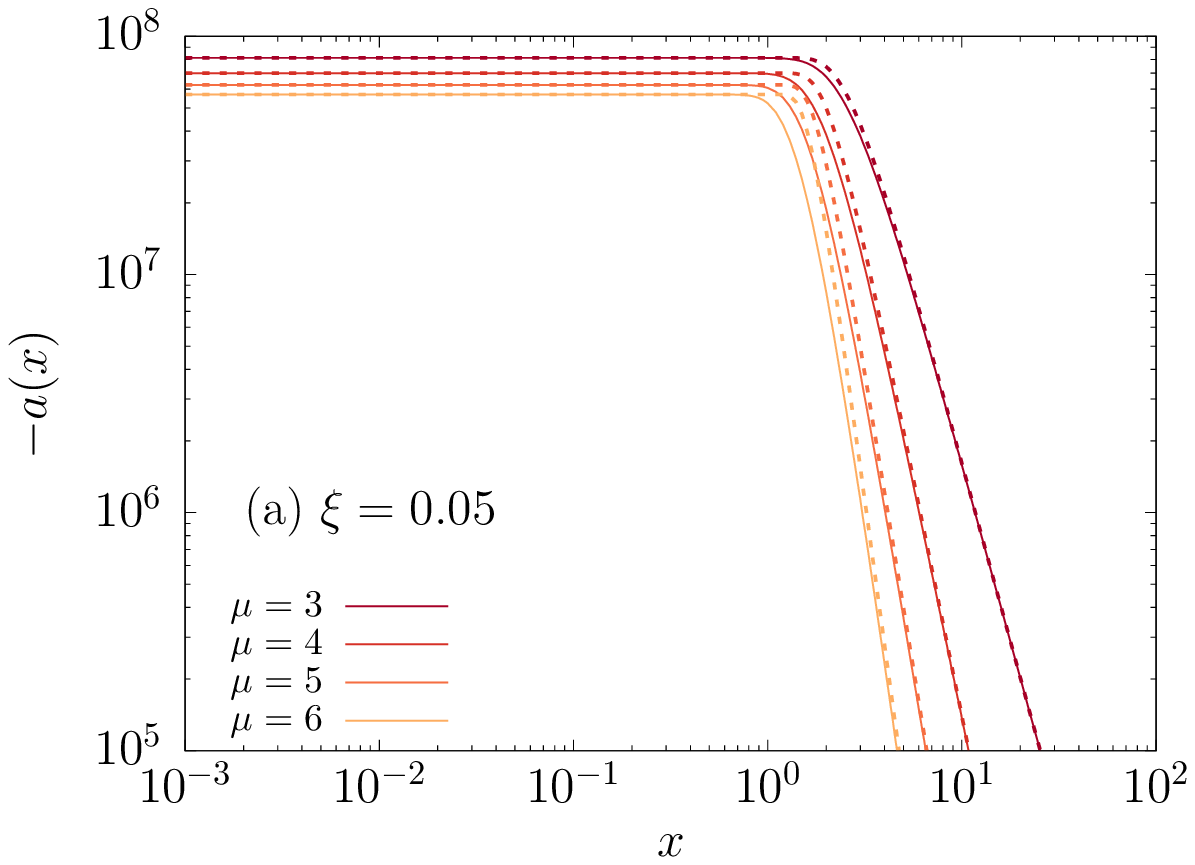}
	\includegraphics[width = 0.49\textwidth]{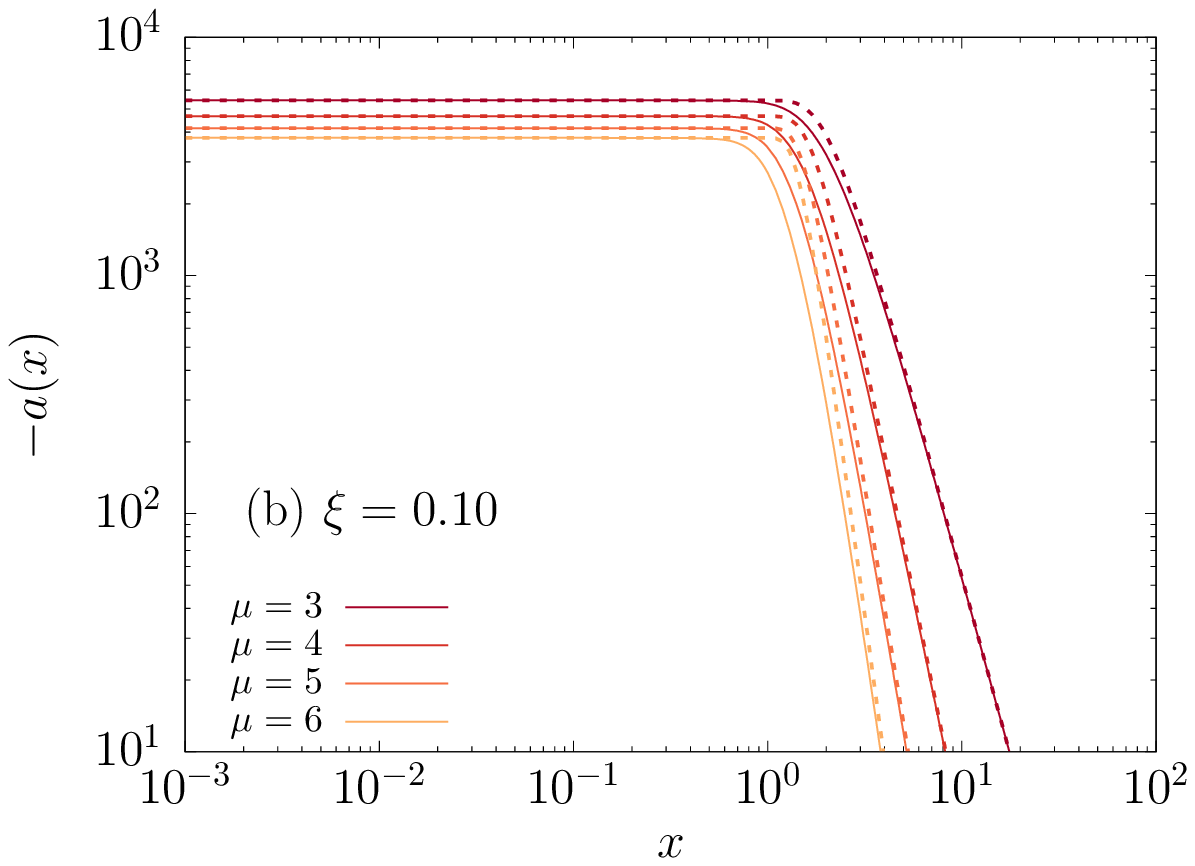}
	\caption{The integrand $a(x)$ of ${\cal A}$, for different values of $\xi$ using the field $v_\mu(x) = - {1}/{(1+x^2)^{\mu/2}}$, with several values of $\mu$. Solid lines represent a direct numerical evaluation of $a(x)$ and the dotted line represents our approximation in Eq. (\ref{eq:ap-a}). \label{fig:compare-a}}
\end{figure}

The next question is how the deepness of the potential affects $\ell_0$  and ${ \cal A}$. 
The calculation of ${\cal A}$ is a bit more challenging. With regard to the integrand of its definition, 
$a(x) = { e}^{v(x)/\xi}g(x) - x - \ell_0$, its value at 0 is $-\ell_0$, which, as we have already seen, is exponentially large. As shown in Fig.~\ref{fig:compare-a}, it is basically constant till some $x_*$, then it decays as a power-law. For large $x$,   assuming that $v(x) \approx - 1/x^\mu$,   with $\mu>1$, we have
\begin{eqnarray}
a(x) \approx { e}^{-\frac{1}{\xi\, x^\mu}} \left( x + \ell_0 + \frac{1}{\xi(\mu-1)x^{\mu-1}} \right) - x - \ell_0 \, .
\end{eqnarray}
The largest terms by far are those proportional to $\ell_0$, and so
\begin{eqnarray}
a(x) \approx \ell_0 \left({ e}^{-\frac{1}{\xi \, x^\mu}} - 1\right).
\label{eq:ap-a}
\end{eqnarray}

\begin{figure}[h!]
	\centering
	\includegraphics[width = 0.49\textwidth]{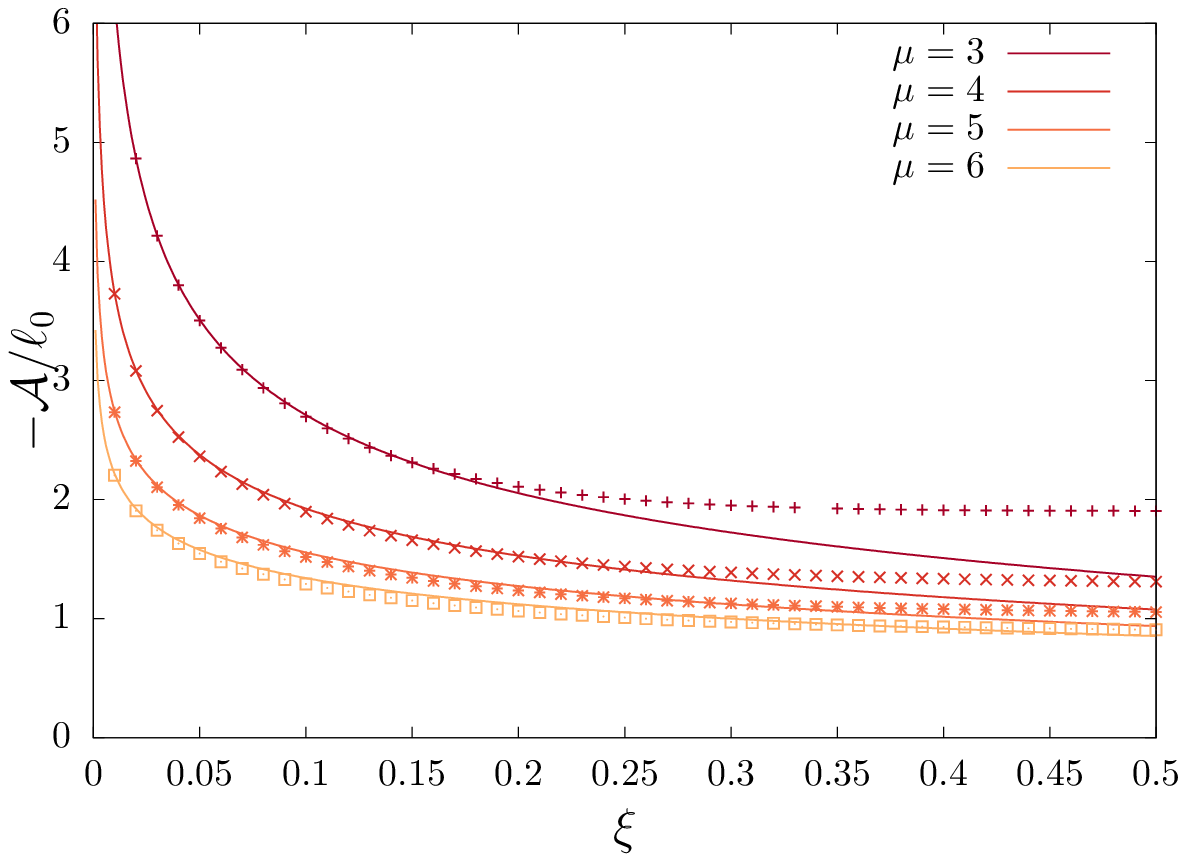}
	\includegraphics[width = 0.49\textwidth]{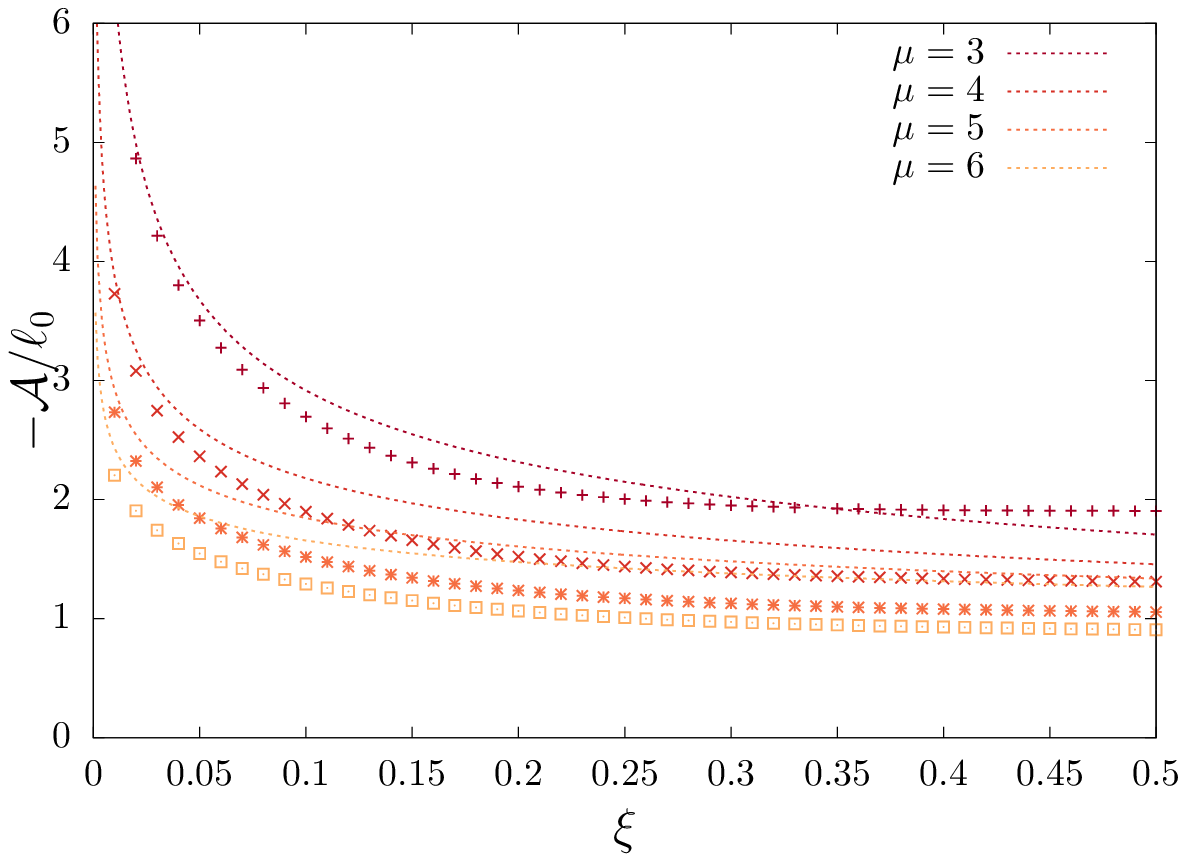}
	\caption{The ratio $\phi = -{\cal A}/\ell_0$ vs. $\xi$,  the scaled temperature of the system, where ${\cal A}$ and $\ell_0$ are given by Eqs.~(\ref{eq:A}) and (\ref{eq:l0}), respectively. We plot the direct solution (points) with the leading order (right panel, dotted lines), Eq.~(\ref{eq:A1}), and the leading order with the first correction (left panel, solid line), Eq.~(\ref{eq:A2}). 
	\label{fig:compare-b}}
	
\end{figure}

We show this approximation for $v(x) = -1/(1+x^2)^{\mu/2}$ 
in Fig.~\ref{fig:compare-a}. Integrating $a(x)$, we find
\begin{eqnarray}
{\cal A} \approx - \frac{\ell_0}{\xi^{1/\mu}} \Gamma \left( \frac{\mu-1}{\mu} \right)\, \label{eq:A1}
\end{eqnarray}
so that ${\cal A}$ is exponentially large and has the opposite sign of $\ell_0$ (see Fig.~\ref{fig:compare-b}). For our example, we can calculate the next correction as well, and
\begin{eqnarray}
a(x) \approx \ell_0 \left\{ {e}^{-\frac{1}{\xi \, x^\mu}} \left( 1 + \frac{\mu}{ 2 \xi x^{\mu+2}} \right) - 1 \right\}
\end{eqnarray}
and
\begin{eqnarray} \label{eq:A2}
{\cal A} \approx - \ell_0  \left\{ \frac{1}{\xi^{1/\mu}} \Gamma \left( \frac{\mu-1}{\mu} \right)- \frac{\xi^{1/\mu}}{2} \Gamma\left( \frac{\mu+1}{\mu} \right) \right\}\,.
\end{eqnarray}

In the matching region \textbf{II}, where $1 \ll x \ll 1/k$,  based on Eq.~(\ref{eq:SMA3}), 
\begin{eqnarray}
\Psi_k^{\bf II}(x) \approx { e}^{v(0)/\xi} \left( 1 - k^2 \bigl(\frac{x^2}{2} + \ell_0 x + {\cal A} \bigr) \right).
\end{eqnarray}
Note that, as long as $x$ is not too large, the last two terms are dominant.

In region {\bf III}, since $x \gg 1$, the $v''(x)$ and $v'^2(x)$ terms are negligible, and therefore, Eq. (\ref{eq:SMA3}) now reads
\begin{eqnarray}
\frac{\partial^2}{\partial x^2}\Psi_k(x)  \sim - k^2 \Psi_k(x).
\end{eqnarray}
Comparing to the region {\bf II} solution, we get
\begin{eqnarray} \label{eq:psiIII}
\Psi_k^{\bf III}(x) \approx { e}^{v(0)/\xi} \left\{ \cos(kx) - k \ell_0 \sin(k(x-\phi)) \right\}\,,
\end{eqnarray}
where 
\begin{equation} \label{eq:phi}
\phi = - {\cal A}/\ell_0 \approx \Gamma(3/4)/\xi^{1/4}\,.
\end{equation}
Let us remark that this derivation is based on 
a Dirac delta function at the origin as initial condition, 
however, 
the shift $\phi$ is expected to be the same for other initial distributions where almost all particles are in the effective region of the potential.  
 
In Eq.~(\ref{eq:psiIII}), since the $k'$s are of order $1/L$, the sin term  is dominant as long as $L \ll { e}^{v(0)/\xi}$. We can now calculate the normalization, 
\begin{eqnarray}
N_k \approx L k^2 \ell_0^2 {e}^{2 v(0)/\xi}.
\end{eqnarray}
Thus, in region {\bf III}, we have
\begin{eqnarray} \label{eq:P3sum}
P^{\bf III}(x,t) = { e}^{v(0)/\xi} \left\{ {\cal I}(x) N_0^{-1} + \sum_{\{ k \}} N_k^{-1} \Psi^{\bf III}_k(x) \, {e}^{-k^2 t} \right\} \, .
\end{eqnarray}
 For large $L$, the spectrum becomes continuous, then the sum over $\{k\}$ transforms into an integral, $\sum_k \to \frac{L}{\pi}\int dk $. 
However, notice that, 
at the same time that $\sqrt{t} \ll L$,  
 it must be $L \ll  {e}^{1/\xi}$. 
 This latter constraint is crucial, as it ensures
 that the Boltzmaniann central part of the PDF (equilibrium state) has almost unit weight relative to the tails (continuum states). 
We achieve this by preventing $L$ from being too large (see {\it bounded domain approach} in~\cite{previous2020}).

Replacing the sum in Eq.~(\ref{eq:P3sum}) and doing the integral yields  
\begin{eqnarray} \label{eq:P3}
P^{\bf III}(x,t) \approx \frac{1}{2 \ell_0} \left\{ 1 - {\rm  erf}\left( \frac{x - \phi}{2 \sqrt{t}} \right) \right\} = \frac{1}{2 \ell_0}\, {\rm erfc}\left( \frac{x - \phi}{2 \sqrt{t}} \right)\,.
\end{eqnarray} 
The shift $\phi$, given by Eq.~(\ref{eq:phi}), in this free diffusion solution, is induced by the well. It delimits the region of the well that has to be overcome to escape.

Our calculation for intermediate-times rests fundamentally on the assumption that very little flux has yet escaped the well. %
We can calculate the time where this assumption breaks down by examining how much probability has flowed from region {\bf I} to region {\bf III}.  Considering a point $x=\ell\sim \phi$, given by Eq.~(\ref{eq:phi}), where regions {\bf I} and {\bf III} overlap, the whole probability in region {\bf III} can be written as
 
\begin{eqnarray}\nonumber  
\int_\ell^\infty P^{\bf III}(x,t) dx &=& \frac{1}{2\ell_0} \left\{ \frac{2 t^{1/2}}{\sqrt{\pi}} \; e^{\displaystyle -\frac{(\ell-\phi)^2}{4t}} + (\phi - \ell)\; {\rm erfc}\left( \frac{\ell - \phi}{2\sqrt{t}} \right) \right\} \\ \label{eq:p3larget}
    &\approx & \frac{t^{1/2}}{\sqrt{\pi} \, \ell_0 } + \frac{\phi- \ell}{2 \ell_0} + O(t^{-1/2}) \,.
\end{eqnarray}
The last line in Eq.~(\ref{eq:p3larget}) is the large-$t$ expansion. 
$\int_\ell^\infty P^{\bf III}(x,t) dx$    is small. 
Then, we conclude that the intermediate-long-time limit holds for times $t$ such that
\begin{eqnarray}
\sqrt{t} \ll \sqrt{\pi} \ell_0 \propto {e}^{-v(0)/\xi} = {e}^{1/\xi} \, ,
\end{eqnarray}
which is the Arrhenius factor~\cite{rednerbook,Kramers,Arrhenius}. 
Let us remark that, unlike Kramers escape problem, which is related to ours, here the potential field is flat at large $x$, and this makes the two problems non-identical.

\begin{figure}[h!]
	\centering
	\includegraphics[width = 0.48\textwidth]{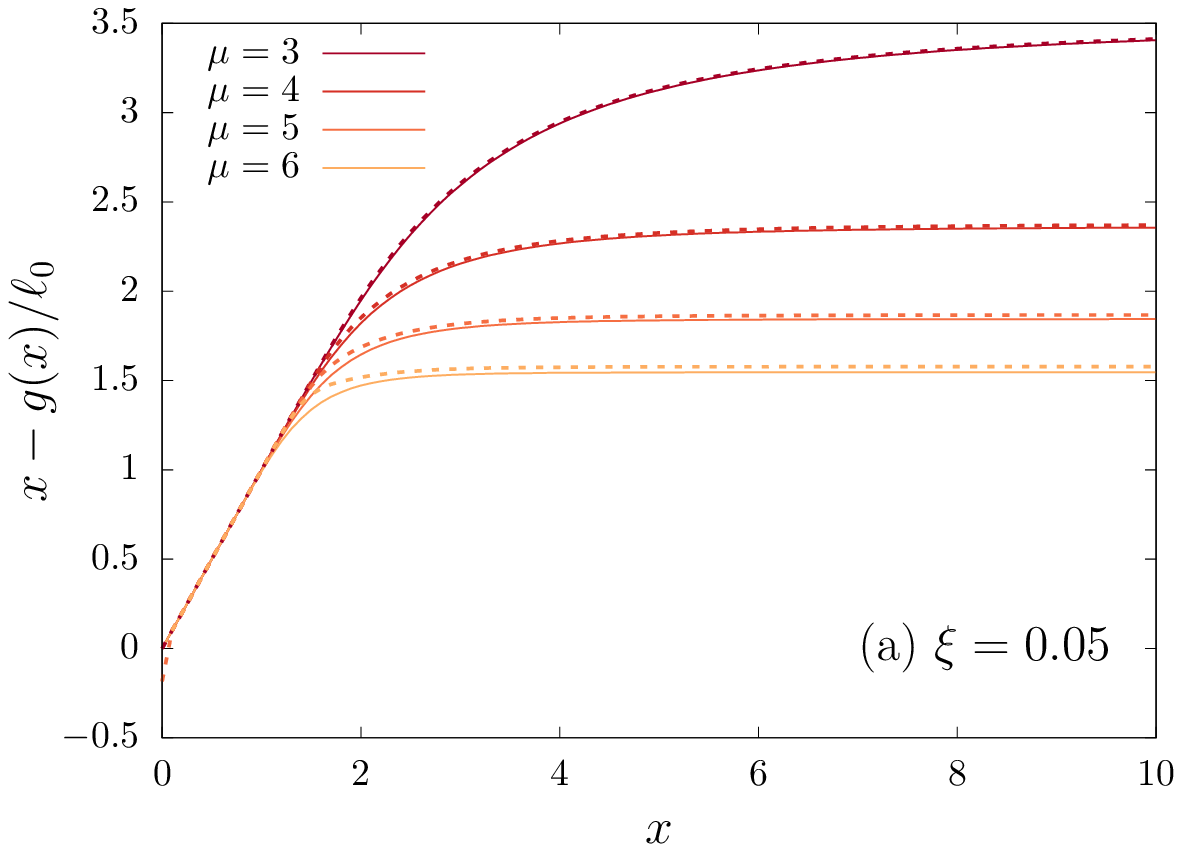}
	\includegraphics[width = 0.48\textwidth]{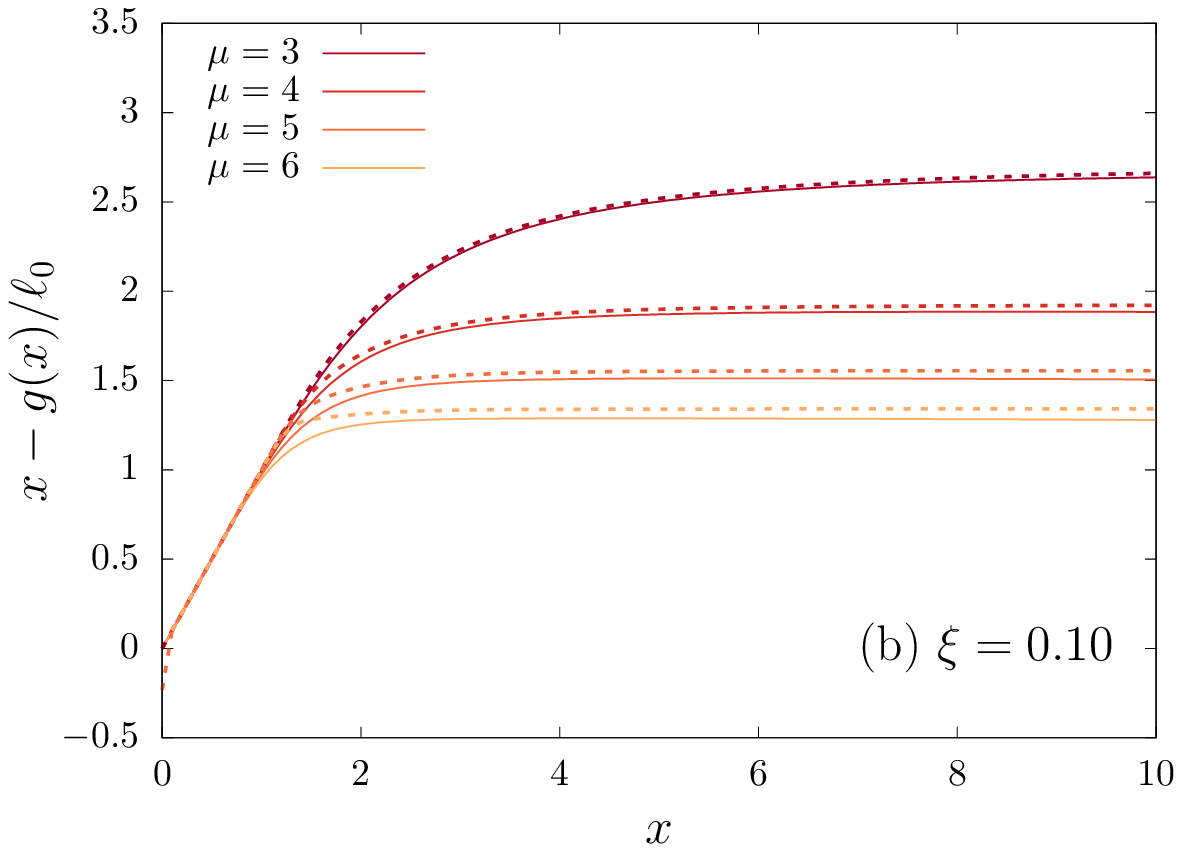}
	\caption{The ratio $x - g(x)/\ell_0$ vs. $x$ calculated  directly (solid line) from Eq.~(\ref{eq:g}), together with our deep well approximation (dotted line), Eq. (\ref{eq:f-deep-well}) for $v_\mu(x) = -1/(1 + x^2)^{\mu/2}$ with several values of $\mu$ and two values of $\xi$.}
	\label{fig:compare-3} 
\end{figure}

To get an approximation for $P(x,t)$ in region {\bf I} we need the small $\xi$ approximation to the function $g(x)$, see Eq.~(\ref{eq:g-ap}). This works similarly to our small $\xi$ approximation for ${\cal A}$. We have
\begin{eqnarray} 
g(x) = \frac{x^2}{2} + \ell_0 x + \int_0^x \left\{ { e}^{v(x_1)/\xi} \left( g_1(x_1) - x - \ell_0 \right) +  { e}^{v(x_1)/\xi} (x_1 + \ell_0) - x_1 - \ell_0  \right\} dx_1 \, .
\end{eqnarray}
In the integral over $x_1$, the first term within the brackets is clearly not exponentially large, and so the only terms proportional to $\ell_0$ are
\begin{eqnarray}
g^{\bf I}(x) \approx \ell_0 \left( x - \int_0^x \left( { e}^{v(y)/\xi} - 1 \right) dy \right) \, .
\end{eqnarray}
For our first standard example, $v(x) = - 1/(1+x^2)^{\mu/2}$, we have
\begin{eqnarray} \label{eq:gI}
g^{\bf I}(x) &\approx &\ell_0 \left\{ x + \int_0^x \left(  { e}^{-\frac{1}{\xi \, y^\mu}} \left( 1 + \frac{\mu}{ 2 \xi y^{\mu+2}} \right) - 1  \right) dy \right\}  \\
&= & \ell_0 \left\{ x - \frac{x}{\mu} E_{1+1/\mu}\left(\frac{1}{\xi x^\mu}\right) + \frac{\xi^{1/\mu}}{2} \Gamma\left( 1 + 1/\mu, \frac{1}{\xi x^\mu} \right) \right\} \, , \label{eq:f-deep-well}
\end{eqnarray}
where $E_n(x)$ is the exponential integral function and $\Gamma(n,x)$ is the incomplete Gamma function, 
outcomes of the calculation of the integral in Eq.~(\ref{eq:gI})  using Wolfram Mathematica \cite{mathematica}.
This is verified in Fig. \ref{fig:compare-3}, where we plot $\bigl(x - g(x)\bigr)/\ell_0$ 
vs. $x$, together with our analytic approximation.

\begin{figure}[h!]
\centering
\includegraphics[width=0.48\textwidth]{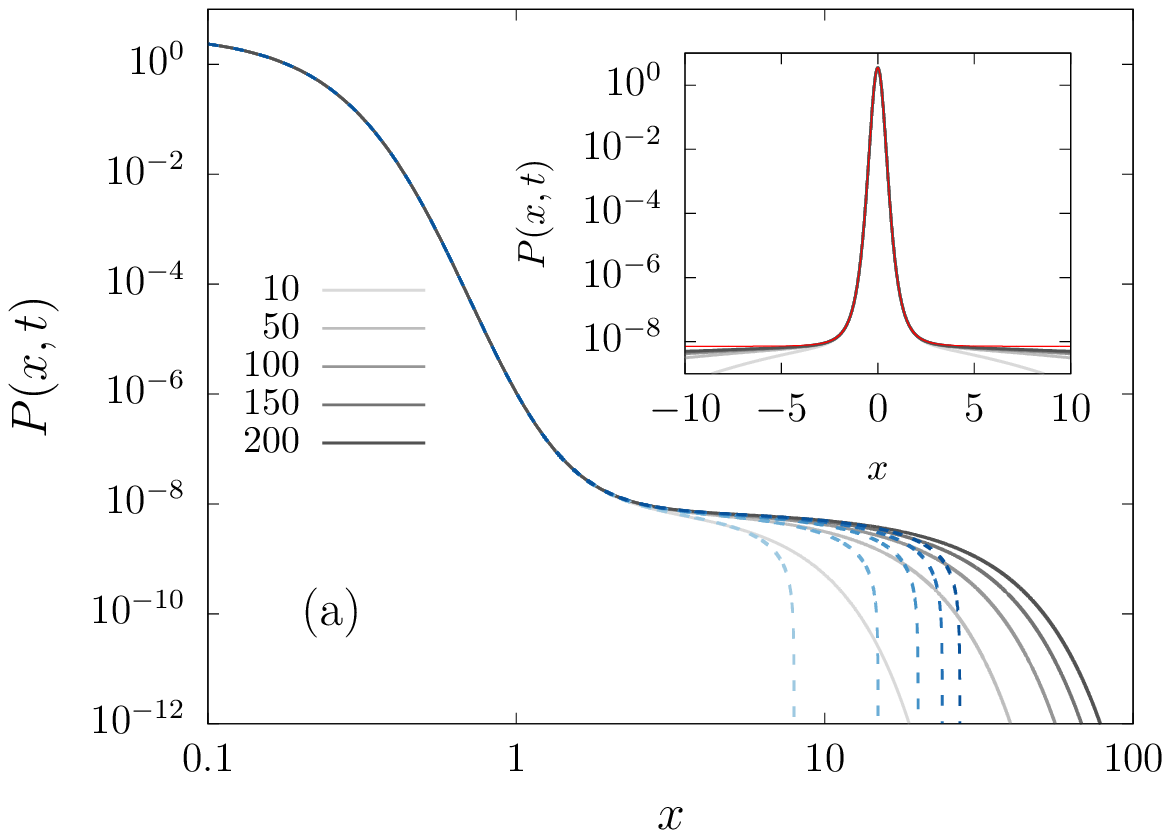}
\includegraphics[width=0.48\textwidth]{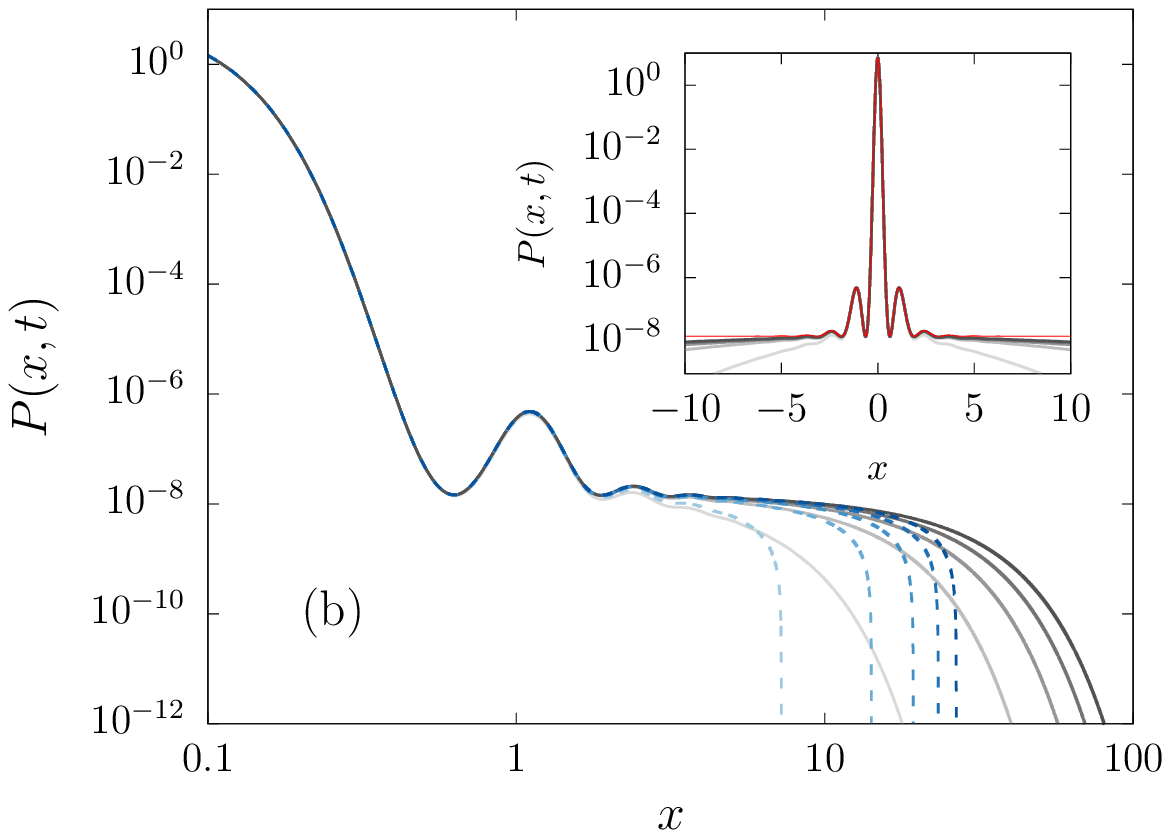}
\includegraphics[width=0.48\textwidth]{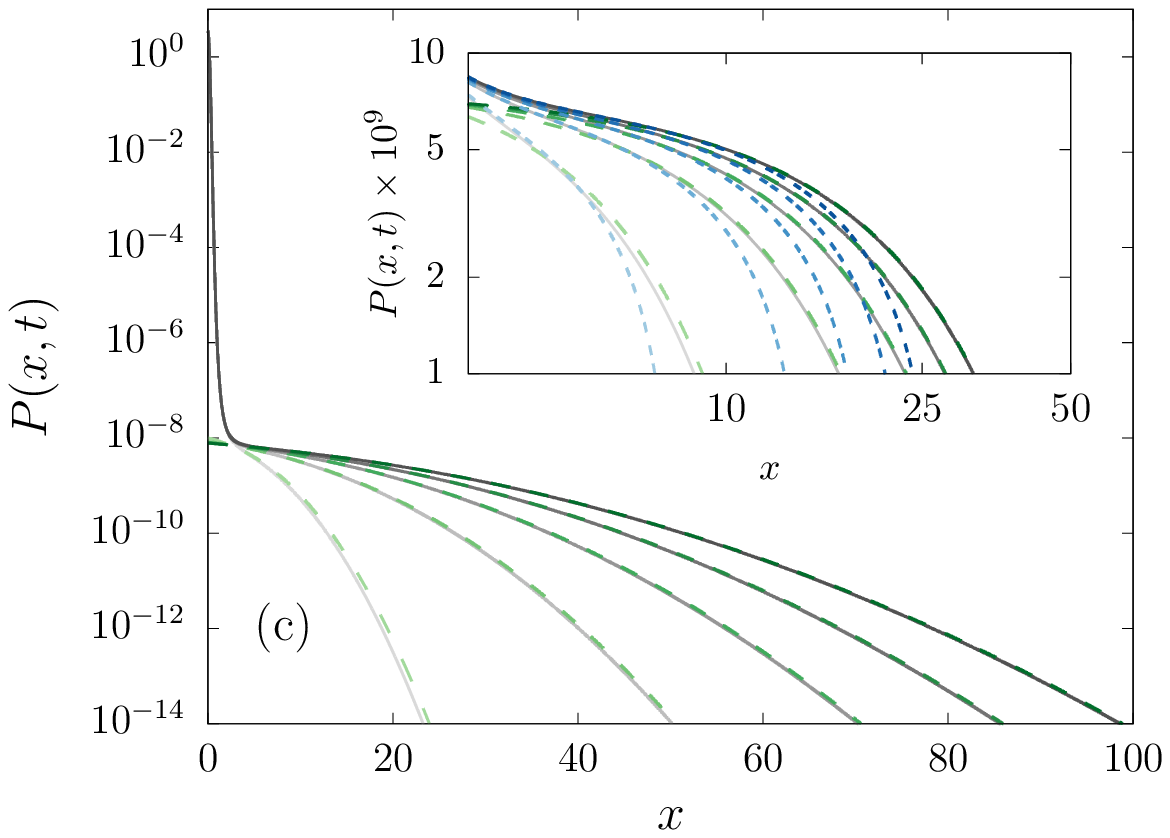}
\includegraphics[width=0.48\textwidth]{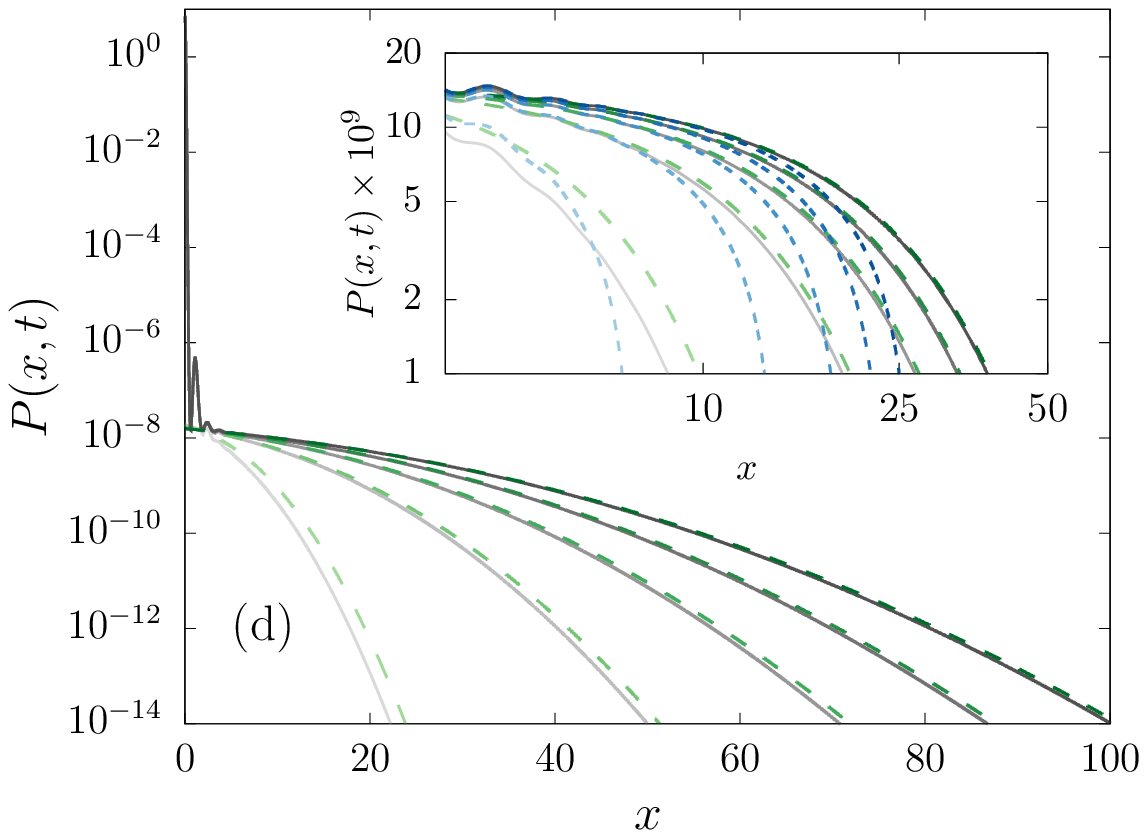}
\caption{
Central region {\bf I} (top row) and tail region {\bf III} (bottom row) of the PDF, for the potentials 
$v_4$ (left column) and $v_{4,5}$ (right column), at different times $t$ (increasing from light to dark color) indicated in the legend. 
PDF from the numerical integration of the FPE (black solid lines) and theoretical predictions (dashed lines): $P^I(x,t)$ (blue short dashed) given by Eq.~(\ref{eq:P1}), and  $P^{III}(x,t)$ (green long dashed) given by (\ref{eq:P3}). 
The upper insets include the Boltzmannian (red dashed) curve for comparison (duplicating Fig.~\ref{fig:pdf}).
The intermediate matching region is amplified in the lower insets. 
}
\label{fig:regions}
\end{figure}

Thus,
\begin{eqnarray}  \label{eq:P1}
P^{\bf I}(x,t) &\approx & \frac{{\cal I}(x)}{2 { e}^{v(0)} \ell_0} \left(1 - g^{\bf I}(x) \int_0^\infty \frac{dk}{\pi \ell_0} { e}^{- k^2 t} \right) \nonumber \\
& \approx & \frac{{ e}^{-v(x)/\xi}}{2 \ell_0} \left( 1 - \frac{1}{\sqrt{\pi t}} \frac{g^{\bf I}(x)}{\ell_0} \right),
\end{eqnarray}
which overlaps, as it must, with the region {\bf III} result. For sufficiently large $t$, Eq.~(\ref{eq:P1})  becomes
\begin{eqnarray}
P^I(x,t) \approx  {e}^{ -v(x)/\xi}/(2 \ell_0) \, , \label{eq:NQE-BG}
\end{eqnarray}
which is time-independent. Notice that, 
as mentioned earlier,  
$2 \ell_0$ serves as an effective partition function, 
which replaces $Z$ in Eq.~(\ref{eq:BG}).
Moreover,   region {\bf I} is where most of the probability is found, and hence this expression captures the behavior of the majority of the particles.

Figure~\ref{fig:regions} summarizes the behavior of the PDF in 
regions {\bf I} (center) and {\bf III} (tail). 
(The upper insets replicate  Fig.~\ref{fig:pdf}, to 
complete the portrait.)
The results from numerical simulations are represented 
by solid lines while the theoretical results  
for regions {\bf I} (short dashed line) and {\bf III} (long dashed lines) are also plotted, in good agreement in the 
respective regions.
In the insets all the curves are plotted together, in a zoom of the matching region.

 \section{Regularization procedure}
 \label{sec:regularization}
 
 Since Eq.~(\ref{eq:NQE-BG}) is time-independent, averages computed when the system is inside the intermediate time-scale epoch outlined in Section~\ref{sec:timesolutions} become almost constant in time. Even though it is not possible to apply the regular BG equilibrium statistics, for systems where $\xi$ is small  we can, through a regularization procedure, obtain time-independent averages, which we refer as NQE averages. 
 We now demonstrate the regularization procedure that allows us to compute these averages.

 The average of a given observable ${\cal O}(x)$ can be obtained using the PDF as
 \begin{eqnarray}
  \langle {\cal O} \rangle &=& \int_{-\infty}^{\infty} {\cal O}(x) \, P(x,t) \, dx  \, .
 \end{eqnarray}
 We can split the integration in two regions $(x,\ell)$ and $(\ell,\infty)$, 
 where $\ell$ is an intermediate length scale 
 ($\ell \sim \phi=-{\cal A}/\ell_0$, defined above), namely,

 \begin{eqnarray}  \label{eq:O13}
  \langle {\cal O} \rangle &\simeq & 2 \int_{0}^{\ell} {\cal O}(x) \, P^{\bf I} (x,t) \, dx + 2 \int_{\ell}^{\infty} {\cal O}(x) \, P^{\bf III} (x,t) \, dx  
  =  \langle {\cal O} \rangle_{\bf I} + \langle {\cal O} \rangle_{\bf III} \,.
 \end{eqnarray}
Recalling that, for intermediate timescales, region {\bf I} concentrates most of the probability and $P^I$ becomes 
nearly time-independent, then Eq.~(\ref{eq:O13}) allows 
to predict the NQE value. 
It is noteworthy that the NQE regime of different observables is related to different timescales.

We will now illustrate this procedure through the computation of the average MSD, defined in Eq.~(\ref{eq:MSDt}), for a system subject
to the potential fields with a power-law decay. 
This is a simple still good example since the MSD 
constitutes a relevant dynamical measure of the spread of the particles around the central potential well, 
as defined in Eq.~(\ref{eq:MSDt}).   

First we calculate $\langle x^2 \rangle_{\bf I}$, 
using  (\ref{eq:P1}). After neglecting the correction containing $g^I(x)$   for $t \gg \ell_0$, we  perform the same trick used in Eq.~(\ref{eq:ll0}) to obtain $\ell_0$, namely, 

\begin{eqnarray} \label{eq:MSD1a} 
&\langle x^2\rangle_{\bf I}  &
 \simeq  
\frac{1}{\ell_0}  \int_{0}^{\ell} x^2 \,
 { e}^{-v(x)/\xi} \left( 1 - \frac{1}{\sqrt{\pi t}} \frac{g^{\bf I}(x)}{\ell_0} \right) \, dx \\  \label{eq:MSD1b}
 & \simeq & 
  \frac{1}{\ell_0} \int_{0}^{\infty} x^2 \,
 \Bigl( { e}^{-v(x)/\xi} - h(x) \Bigr) dx -
 \frac{1}{\ell_0} \int_{\ell}^{\infty} x^2 \,
 \Bigl( { e}^{-v(x)/\xi} - h(x) \Bigr) dx   
 +\frac{1}{\ell_0}\int_0^\ell x^2 h(x) dx
  \\  
 &\simeq&  \label{eq:MDS1final}
  \frac{1}{\ell_0} \int_{0}^{\infty} x^2 \,
 \Bigl( { e}^{-v(x)/\xi} - h(x) \Bigr) dx \,.
\end{eqnarray}

Differently from the case of Eq.~(\ref{eq:ll0}), 
for integral convergence, we must set 

\begin{equation} \label{eq:h}
h(x)= \sum_{k=0}^K (-1)^k (v(x)/\xi)^k/k! \,,
\end{equation}
where we sum up to the integer $K$ defined, for the specific observable, as the minimal value to ensure that the integral converges. In the specific case of $x^2$, in Eq.~(\ref{eq:MSD1a}), 
we have $K = \lfloor 3/\mu \rfloor $, and for a general $x^n$ we have $K = \lfloor (n+1)/\mu \rfloor $.  The reasoning behind this technique goes beyond merely creating a converging integral, for small values of $\xi$ we have $h(x) \ll e^{-v(x)/\xi}$ for the range $(x,\ell)$, therefore we are able to {\it cure} the diverging contribution from the tail whilst maintaining a very accurate result overall.

In particular, for $\mu>3$, we have $h(x)=1$. 
The first term in Eq.~(\ref{eq:MSD1b}) 
is the only time-independent term, related to the standard BG probability. 
Recall that, from Eq.~(\ref{eq:l0}),
$
\ell_0 \equiv \int_0^\infty \left( { e}^{-v(x)/\xi} -1 \right)dx$, of order ${ e}^{1/\xi}$. 
The second term  scales as $1/\xi$, so that it becomes increasingly negligible 
compared to $\ell_0$. 
The same occurs for the last term in Eq.~(\ref{eq:MSD1b}), which for $h(x)=1$ becomes $\ell^3/(3\ell_0)$.

Now, we calculate $\langle x^2 \rangle_{\bf III}$, 
using  (\ref{eq:P3}). In this case it is  possible to perform the integral  exactly to obtain

\begin{eqnarray}
   \langle x^2 \rangle_{\bf III} &=& 
    \frac{1}{\ell_0} \int_{\ell}^{\infty} x^2 \,  {\rm erfc}\left( \frac{x - \phi}{2 \sqrt{t}} \right) 
\, dx \\ \noindent
   &=&  
   \frac{1}{3\sqrt{\pi}\, \ell_0}
  \left( 2\sqrt{t} \,(\phi^2 +\ell\phi +\ell^2 + 4t) 
  \,{e}^{-\frac{(\ell-\phi)^2}{4t} }+
 \sqrt{\pi} \,(\phi^3 -\ell^3 + 6\phi t) \,{ \rm erfc} \left[ \frac{\ell-\phi}{2\sqrt{t}}\right]
  \right)  \\ \label{eq:x2-III}
  & \approx & \frac{8 t^{3/2}}{3 \ell_0 \sqrt{\pi}} + 2 \phi t + \frac{2 \phi^2 t^{1/2}}{\ell_0 \sqrt{\pi}} + \frac{\phi^3 - \ell^3}{3 \ell_0} + O(t^{-1/2}) \, ,
\end{eqnarray}
where the last member of the equation is obtained from the large-$t$ expansion. 

Putting this all together, we write, up to the first correction for large time,
\begin{eqnarray}
\langle x^2 (t) \rangle \simeq \langle x^2 \rangle_{{\bf I}} 
+ \langle x^2 \rangle_{{\bf III}} 
\simeq \langle x^2 \rangle_{{\bf I}} +  
8 t^{3/2}/[3\sqrt{\pi}l_0 ] \, . \label{eq:MSD-approx}
\end{eqnarray}
The average will become almost time-independent for time-scales $t$ such that
\begin{eqnarray}
t \ll \frac{ \left( 3 \ell_0 \right)^{2/3}}{4 \pi^{1/3}} \propto e^{-2v(0)/3\xi} \,,
\end{eqnarray}
which is also related to the Arrhenius factor~\cite{rednerbook, Kramers,Arrhenius}. The time-dependent contribution will be negligible, and for large times, we can estimate the departure times.
Then,  the NQE average is estimated (when $\mu>3$) as
\begin{eqnarray} \label{eq:NQEteo} \langle  x^2 \rangle_{\rm NQE} \simeq \langle  x^2 \rangle_{\bf I} \simeq
 \frac{1}{\ell_0}\int_0^\ell x^2 e^{-v(x)/\xi} dx \approx \frac{\int_0^\infty x^2 \left(  e^{-v(x)/\xi} -1 \right) dx}{\int_0^\infty \left( e^{-v(x)/\xi} - 1 \right) dx}\,.
\end{eqnarray}
The performance of these approximations can be appreciated in Fig. \ref{fig:MSDvst}, 
for different values of $\xi$. The smaller is $\xi$, the longer is the lifetime of the NQE regime and the better works the theoretical prediction for the NQE level, given by Eq.~(\ref{eq:NQEteo}). 
The figure also exhibits the improvement of the  theory for the NQE  with respect to the harmonic approximation of the potential well (dotted lines).

 For a general observable, and potentials decaying faster that $1/x$ ($\mu>1$), the NQE average is 
 \begin{eqnarray} \label{eq:theory}
 \langle  {\cal O}(x) \rangle_{\rm NQE} = \frac{\int_0^\infty {\cal O}(x) \left(  e^{-v(x)/\xi} -h(x) \right) dx}{\int_0^\infty \left( e^{-v(x)/\xi} - 1 \right) dx} \, ,
 \end{eqnarray}
 where $h(x)$ is defined as Eq.~(\ref{eq:h}), ensuring convergence. 
  Therefore, it is determined by the observable and by the potential field~\cite{previous2020}. 
  For potentials with $0< \mu \le 1$,  $\ell_0$ must be modified, hence the 
  denominator in Eq.~(\ref{eq:theory}) becomes 
 
  \begin{eqnarray}
  \ell_0=\int_0^\infty \left( e^{-v(x)/\xi} - l(x) \right) dx
  \end{eqnarray}
with

\begin{equation} \label{eq:l}
l(x)= \sum_{k=0}^{K'} (-1)^k (v(x)/\xi)^k/k! \,,
\end{equation} 
where $K' = \lfloor 1/\mu \rfloor $.

 \section{Final Remarks}
 \label{sec:final}
 
We presented the solution of the FPE (\ref{eq:FPEscaled}) for asymptotically flat  potentials with  a deep well at the origin, using an eigenfunction expansion. 
In such potential fields, long-lived NQE states emerge, 
as  heuristically shown in our previous work~\cite{previous2020}. %
The non-confinement of the potential makes the standard partition function divergent, hampering its direct application. Nevertheless, a regularization procedure is still
possible, allowing one to calculate quantities in the NQE states
along the lines of the recipes of statistical mechanics 
(see Eq.~(\ref{eq:theory})). 

The spectrum of eigenvalues is continuous like that of a free particle, still the Boltzmann measure is preserved for intermediate times 
such that $\sqrt{t} \ll \sqrt{\pi} \ell_0 \propto  {e}^{1/\xi}$ which is the Arrhenius time.  
In such case, according to Eqs.~(\ref{eq:P3}) and (\ref{eq:NQE-BG}), 
the approximate solution we found can be summarized as 

\begin{eqnarray} \label{eq:summary}
P(x,t) = \left\{ \begin{array}{ll}
 \frac{1}{2 \ell_0}\, {e}^{ -v(x)/\xi},  & \mbox{region {\bf I}},\\
 \frac{1}{2 \ell_0}\, {\rm erfc}\left( \frac{x - \phi}{2 \sqrt{t}} \right), &\mbox{region {\bf III}},\\
\end{array} 
\right.
\end{eqnarray}
where region {\bf I} corresponds to the central part of the PDF
and region {\bf III} to the tails, while region {\bf II} is 
where both solutions overlap, as can be seen in Fig.~\ref{fig:regions};
moreover,  $2 \ell_0$, defined in Eq.~(\ref{eq:l0}), 
is a lengthscale that plays the role of an effective partition function, 
and the shift $\phi$ can be  estimated through Eq.~(\ref{eq:phi}).
Region {\bf I} concentrates most of the probability, out of fluctuations 
in region {\bf III}, where the erfc function acts as an effective cutoff 
blocking free diffusion.
The shift is related to the region of the well that has to be overcome 
to escape.  
To see this note that $l_0$ is large but $e^{- v(0)/\xi} = e^{1/\xi}$ is similarly large (while the erfc is of order one or less) hence the small $x$ solution in region {\bf I} is exponentially  exceeding the solution in region {\bf III}.
Then, the shift $\phi$ decreases with increasing relative temperature $\xi$, as 
can be seen in  Fig.~\ref{fig:compare-b}. 
Eq.~(\ref{eq:summary}) shows how nearly time-independent solutions can emerge. 
They last exponentially long times, for sufficiently low temperatures, and can be associated to the NQE regime.

The physics of non-normalizable states has been 
the object of extensive studies  within infinite ergodic theory~\cite{nonergodic,akimoto},  in situations different from what we consider here. 
For example, in cases 
where the particle escapes and returns to the well many times,
 the density in region {\bf I} decays in time, while in our case
it remains nearly constant. 
 Notice also how our approach differs from the calculation of observables using 
scaling properties of the PDF~\cite{hanggi}, in the limit of $t \to \infty$. Here we avoid the limit of infinite time considering the upper bound, escape time, $e^{1/\xi}$, which allows us to isolate the dominant Boltzmann-like behavior at the center.

Furthermore, let us mention that preliminary results indicate that some form of ergodicity holds in NQE states. If we restrict the time integration over trajectories to the interval where the observable is in its plateau, i.e., after a transient and for times shorter than Arrhenius time, we will obtain the same result as the NQE ensemble average in Eq.~(\ref{eq:theory}).
 
 NQE states can emerge in a wider range of systems and observables~\cite{previous2020}, 
beyond the MSD used here in our examples, as long as there is a clear separation of lengthscales between the effective well and the long tail behavior. 
This indicates that the method is rather robust, in the sense that it is not restricted to potentials with a single well. 
For sufficiently separated wells, more than one plateau can emerge, in that case, the theory predicts the last one, 
before particles escape the full region where forces effective.

The theory will work also in higher dimensions, though the details should be part of a separate study.
The investigation of the fractional Fokker-Planck equation for
anomalous dynamics, as well as generalized Langevin equations with memory would be 
interesting extensions of the present work.

\mbox{}\\[3mm]
{\bf Acknowledgements:}
D.K. and E.B. acknowledge the
support of Israel Science Foundation's grant 1898/17.
C.A. and L.D. acknowledges partial financial support from
Conselho Nacional de Desenvolvimento Cient\'{\i}fico
e Tecnol\'ogico (CNPq), and Funda\c c\~ao de Amparo \`a Pesquisa do Estado do Rio de Janeiro (FAPERJ), 
and Coordena\c c\~ ao de Aperfei\c coamento de Pessoal de N\' ivel Superior - Brazil (CAPES) - Finance Code 001.


\begin{thebibliography}{99}



\bibitem{previous2020} 
L. Defaveri, C. Anteneodo, D.A. Kessler, E. Barkai, 
{\it Regularized Boltzmann-Gibbs statistics for a Brownian particle in a nonconfining field},  
Phys. Rev. Research {\bf 2} (4), 043088 (2020).


\bibitem{Fermi}
E. Fermi, 
{\it Uber die Wahrscheinlichkeit der Quantenzustande},
  Zeitschrift fur Physik  {\bf  26}, 54 (1924).

 
\bibitem{Plastino}
A. Plastino, M. C. Rocca, and G. L. Ferri, 
{\it Resolving the partition function's paradox of theHydrogen atom}, 
  Physica A  {\bf 534}, 15 (2019).


 
\bibitem{freezing2020}
S. Sabhapandit, and S. N. Majumdar, 
{\it Freezing Transition in the Barrier Crossing Rate of a Diffusing Particle},
 Phys. Rev. Lett. {\bf 125}, 200601 (2020).

\bibitem{logpot2011}
A. Dechant, E. Lutz, E. Barkai, D. A. Kessler, 
{\it Solution of the Fokker-Planck Equation with a Logarithmic Potential}, 
Journal of Statistical Physics {\bf 145},  1524–1545 (2011).



\bibitem{vankampen}
N. G. van Kampen, Stochastic Processes in Physics and Chemistry (North-Holland, Amsterdam, New York, 1981). 
  
 
\bibitem{Risken}
 H. Risken,  {\it The Fokker-Planck Equation}, 
In Springer Series in Synergetics: Vol. 18. 
(Springer, Berlin, 1989).

 
\bibitem{rednerbook}
S. Redner, {\it A Guide to First-Passage Processes} (Cambridge University Press, Cambridge, 2001).
 
 \bibitem{Kramers}
 P. H\"anggi, P. Talkner, and M. Borkovec, 
 {\it Reaction-rate theory: fifty years after Kramers}, 
 Rev. Mod. Phys.  {\bf 62}, 251 (1990).
 
 \bibitem{Arrhenius}
 S. Arrhenius, Z. Phys. Chem. (Leipzig) {\bf 4}, 226 (1889).
 
\bibitem{AghionPRL}
 E. Aghion, D. A. Kessler, and E. Barkai, 
{\it From NonNormalizable Boltzmann-Gibbs Statistics to Infinite-Ergodic Theory}, 
Phys. Rev. Lett. 122, 010601 (2019).


\bibitem{Aghion2020}
E. Aghion, D. A. Kessler, E. Barkai,
{\it Infinite ergodic theory meets Boltzmann statistics}, Chaos, Solitons \& Fractals, {\bf 138}, 109890 (2020).


 

\bibitem{sander1998}
D. Kessler, Z. Ner and L. Sander, 
Phys. Rev. E {\bf 58}, 107–114 (1998).

\bibitem{derrida1997}
E. Brunet and B. Derrida, 
Phys. Rev. E {\bf 56}, 2597–2604, (1997).


\bibitem{Montroll} 
J. E. Mayer and E. Montroll, 
{\it Molecular distribution},  J. Chem. Phys.  {\bf 9}, 2  (1941). 

\bibitem{mathematica}\url{https://www.wolfram.com/mathematica/}

  

\bibitem{nonergodic}
 J. Aaronson, {\it Introduction to infinite ergodic theory}, Mathematical Surveys and Monographs
Vol. 50 (1997).

\bibitem{akimoto}
T. Akimoto, E. Barkai, 
{\it Aging generates regular motions
in  weakly chaotic systems},  Phys. Rev. E. {\bf 87}, 032915 (2013).


\bibitem{hanggi}
A. Rebenshtok, S. Denisov, P. H\"anggi, E. Barkai,
{\it  Non-Normalizable Densities in Strong Anomalous Diffusion: Beyond the Central Limit Theorem}, 
Phys. Rev. Lett.  {\bf 112}, 110601 (2014).

 

\end{thebibliography}
\end{document}